\PassOptionsToPackage{hyphens}{url}
\documentclass[sigconf,nonacm]{acmart}

\usepackage{amsmath, amstext}
\usepackage{graphicx}
\usepackage{array, tabularx, threeparttable}
\usepackage{subcaption}
\usepackage{tikz}
\usepackage{pdfrender}
\usepackage{pgfplots}
\pgfplotsset{compat=1.17}
\usepackage{xcolor}
\usepackage{balance}
\usepackage{booktabs}
\usepackage{multirow}
\usepackage{colortbl}
\usepackage{tcolorbox}
\tcbuselibrary{skins}

\usepackage{listings}
\usepackage{makecell}
\usepackage{enumitem}
\usepackage[normalem]{ulem}

\usepackage{float}
\usepackage[algoruled, linesnumbered, noend]{algorithm2e}
\usepackage{xspace}
\usepackage{mdframed}
\usepackage{pifont}
\usepackage{latexsym}
\usepackage{xstring}
\definecolor{linkblue}{HTML}{1F4E79}
\hypersetup{breaklinks=true,colorlinks=true,citecolor=linkblue,linkcolor=linkblue,
            urlcolor=linkblue,filecolor=linkblue}

\definecolor{bblue}{HTML}{4F81BD}
\definecolor{rred}{HTML}{C0504D}
\definecolor{ggreen}{HTML}{9BBB59}
\definecolor{ppurple}{HTML}{9F4C7C}
\definecolor{dkgreen}{rgb}{0,0.6,0}
\definecolor{gray}{rgb}{0.5,0.5,0.5}
\definecolor{mauve}{rgb}{0.58,0,0.82}
\definecolor{darkgray}{rgb}{0.66, 0.66, 0.66}
\definecolor{myred}{HTML}{ebc3b7}
\definecolor{myblue}{HTML}{bfebf3}
\definecolor{mygreen}{HTML}{daebc3}
\definecolor{myyellow}{HTML}{fff4c0}
\definecolor{myorange}{HTML}{ffc58a}
\definecolor{comment-red}{rgb}{0.8,0,0}
\definecolor{gray}{RGB}{136,136,136}

\SetCommentSty{algocommfont}

\definecolor{backcolour}{rgb}{0.95,0.95,0.92}
\definecolor{codegreen}{rgb}{0,0.6,0}
\definecolor{lightlight-gray}{gray}{0.97}

\definecolor{g1}{rgb}{1, 1, 1}
\definecolor{g2}{HTML}{FDF0E2}
\newcommand{\tablestyle}{\footnotesize\setlength{\tabcolsep}{4pt}\renewcommand{\arraystretch}{1.0}}

\definecolor{g3}{HTML}{FFCCCC}
\definecolor{g4}{HTML}{CCFFCC}

\definecolor{dkgreen}{rgb}{0,0.6,0}

\definecolor{rowgray}{gray}{0.96}

\global\mdfdefinestyle{rtboxstyle}{
    linecolor=black,
    backgroundcolor=lightlight-gray,
    leftmargin=0cm, rightmargin=0cm,
    skipabove=4pt, skipbelow=4pt,
    linewidth=0.4pt,
    innertopmargin=3pt,
    innerbottommargin=3pt,
    innerrightmargin=6pt,
    innerleftmargin=6pt,
    roundcorner=2
}

\newcolumntype{L}[1]{>{\raggedright\arraybackslash}p{#1}}
\newcolumntype{C}[1]{>{\centering\arraybackslash}p{#1}}

\newcommand{\CodeIn}[1]{\lstinline{#1}}
\renewcommand{\CodeIn}[1]{\lstinline[basicstyle=\ttfamily]{#1}}

\renewcommand{\paragraph}[1]{\smallskip\noindent\textbf{#1}\quad}
\newcommand{\PP}[1]{\vspace{2px}\noindent{\bf #1}}
\newcommand{\PQ}[1]{\vspace{2px}\noindent\emph{\uline{#1}}}

\newcommand{\PPnew}[1]{\vspace{2px}\noindent{\bfseries #1}\hspace{0.35em}}

\newcommand{\secref}[1]{\S\ref{#1}}

\newcommand{\Sys}{\mbox{\textsc{Encore}}\xspace}
\newcommand{\sys}{\mbox{\textsc{Encore}}\xspace}

\newcommand{\sysCQ}{\mbox{\textsc{Encore}-CodeQL}\xspace}
\newcommand{\sysEX}{\mbox{\textsc{Encore}-Literal}\xspace}

\newcommand{\provg}{\textsc{ProvG}\xspace}
\newcommand{\provgs}{\textsc{ProvG}\xspace}
\newcommand{\poirot}{\textsc{Poirot}\xspace}

\newcommand{\cve}{CVE\xspace}
\newcommand{\cves}{CVEs\xspace}
\newcommand{\poc}{PoC\xspace}
\newcommand{\pocs}{PoCs\xspace}

\newcommand{\ldeg}{log-derived event graph\xspace}

\newcommand{\evtg}{$G_{\mathrm{evt}}$\xspace}

\newcommand{\qg}{$G_{tem}$\xspace}

\definecolor{resback}{HTML}{EEF1F4}
\newtcolorbox{resbox}{enhanced, frame hidden, colback=resback, boxrule=0pt, arc=2pt,
  left=5pt, right=5pt, top=3pt, bottom=3pt,
  before skip=5pt, after skip=5pt}

\DeclareCaptionFont{boldmath}{\boldmath}
\newif\ifdraft \draftfalse
\ifdraft
  \newcommand{\expt}[2]{{\color{red}\textbf{[#1: EXPERIMENT]}~#2}}
  \newcommand{\styl}[2]{{\color{blue}\textbf{[#1: STYLE]}~#2}}
  \newcommand{\subm}[2]{{\color{purple}\textbf{[#1: SUBMISSION]}~#2}}
  \newcommand{\todo}[1]{{\color{red}[TODO: #1]}}
\else
  \newcommand{\expt}[2]{} \newcommand{\styl}[2]{}
  \newcommand{\subm}[2]{} \newcommand{\todo}[1]{}
\fi

\newcommand{\rDarpaPoolPairs}{40{,}000\xspace}

\newcommand{\rEncDatasets}{8\xspace}

\newcommand{\rEncPDatasets}{4\xspace}

\newcommand{\rMatchScrMax}{0.995\xspace}
\newcommand{\rMatchScrMin}{0.946\xspace}

\newcommand{\rPairsNeg}{512\xspace}
\newcommand{\rPairsPos}{512\xspace}

\newcommand{\rScreenPruneMax}{91\xspace}

\newcommand{\rAbInDet}{42\xspace}
\newcommand{\rAbInEv}{10\xspace}
\newcommand{\rAbVpDet}{10\xspace}

\newcommand{\rAcCaptures}{30\xspace}
\newcommand{\rAcCmdCves}{4\xspace}
\newcommand{\rAcCmds}{42\xspace}

\newcommand{\rAcFlows}{2{,}852\xspace}
\newcommand{\rAcJoined}{2{,}850\xspace}
\newcommand{\rAcMaxWorkers}{12\xspace}

\newcommand{\rAhFixedCves}{8\xspace}
\newcommand{\rAhFixedFound}{59\xspace}
\newcommand{\rAhFixedShared}{58\xspace}
\newcommand{\rAhFixedTotal}{64\xspace}

\newcommand{\rAhMaxSim}{0.42\xspace}

\newcommand{\rAlcCves}{10\xspace}
\newcommand{\rAlcFound}{10\xspace}
\newcommand{\rAlcWrong}{0\xspace}

\newcommand{\rAppEpFam}{23\xspace}
\newcommand{\rAppEpHours}{96\xspace}

\newcommand{\rAppWebFam}{43\xspace}

\newcommand{\rApxFixed}{19\xspace}

\newcommand{\rApxLabFixed}{7\xspace}

\newcommand{\rAsMax}{8\xspace}

\newcommand{\rAsMin}{6\xspace}
\newcommand{\rAsPerCve}{8\xspace}

\newcommand{\rAtEvents}{108{,}620\xspace}

\newcommand{\rAtHosts}{2\xspace}

\newcommand{\rBcCqOwn}{28\xspace}

\newcommand{\rBcNvOwn}{23\xspace}

\newcommand{\rBcStOwn}{33\xspace}

\newcommand{\rBdPocMax}{259\xspace}
\newcommand{\rBdPocMedian}{62\xspace}
\newcommand{\rBdPocMin}{14\xspace}

\newcommand{\rBhCveRules}{50\xspace}

\newcommand{\rBtwoRequests}{50{,}000\xspace}

\newcommand{\rCapFound}{38\xspace}
\newcommand{\rCapThird}{28\xspace}
\newcommand{\rCapThirdFound}{26\xspace}
\newcommand{\rCapTotal}{40\xspace}

\newcommand{\rCcEncFp}{2\xspace}
\newcommand{\rCcEncPairs}{854\xspace}
\newcommand{\rCcFold}{10\xspace}
\newcommand{\rCcOtherRulePairs}{18\xspace}
\newcommand{\rCcOtherRules}{11\xspace}
\newcommand{\rCcOutsideFp}{3\xspace}
\newcommand{\rCcRuleAlerts}{166\xspace}
\newcommand{\rCcRuleFp}{20\xspace}

\newcommand{\rCcTopRuleAlerts}{120\xspace}
\newcommand{\rCcTopRulePairs}{2\xspace}
\newcommand{\rClDiff}{19\xspace}
\newcommand{\rClDiffFound}{18\xspace}
\newcommand{\rClLogged}{26\xspace}

\newcommand{\rCompLocalOnly}{4\xspace}
\newcommand{\rCompPlainLost}{2\xspace}

\newcommand{\rCompileMax}{42\xspace}
\newcommand{\rCompileMedian}{0.8\xspace}
\newcommand{\rCorpusLab}{12\xspace}

\newcommand{\rCorpusSplunk}{33\xspace}
\newcommand{\rCorpusTotal}{45\xspace}
\newcommand{\rCostAllAligns}{16{,}321\xspace}
\newcommand{\rCostAllMin}{92\xspace}

\newcommand{\rCostPerAlignMs}{338\xspace}

\newcommand{\rCqContentEv}{5\xspace}

\newcommand{\rCqCsicUnits}{72\xspace}
\newcommand{\rCqDet}{34\xspace}

\newcommand{\rCqLtContent}{676\xspace}

\newcommand{\rCqRungEv}{5\xspace}
\newcommand{\rCqRungGap}{7\xspace}
\newcommand{\rCqRungGapHost}{4\xspace}
\newcommand{\rCqRungGapPay}{3\xspace}

\newcommand{\rCqRungWrongAll}{8\xspace}
\newcommand{\rCqWrong}{14\xspace}

\newcommand{\rCtCves}{6\xspace}

\newcommand{\rCtExpEv}{6\xspace}
\newcommand{\rCtReqs}{140\xspace}

\newcommand{\rEvCthrHi}{12\xspace}
\newcommand{\rEvCthrLo}{2\xspace}
\newcommand{\rEvCves}{10\xspace}

\newcommand{\rEvEncKept}{105\xspace}

\newcommand{\rEvEncTotal}{110\xspace}
\newcommand{\rEvModes}{11\xspace}

\newcommand{\rEvPayKept}{72\xspace}
\newcommand{\rEvPayTotal}{77\xspace}
\newcommand{\rEvPgControls}{5\xspace}

\newcommand{\rEvPgWrongPairs}{45\xspace}
\newcommand{\rEvPoiControls}{9\xspace}

\newcommand{\rExSubDet}{42\xspace}
\newcommand{\rExSubTotal}{44\xspace}

\newcommand{\rExtCsicRequests}{36{,}000\xspace}

\newcommand{\rExtCsicWindowSize}{500\xspace}
\newcommand{\rExtCsicWindows}{72\xspace}

\newcommand{\rFidCves}{10\xspace}
\newcommand{\rFidNamed}{11\xspace}
\newcommand{\rFidSent}{12\xspace}
\newcommand{\rFieldCross}{5\xspace}
\newcommand{\rFieldDet}{43\xspace}

\newcommand{\rHnRequests}{1{,}957\xspace}

\newcommand{\rHostOnly}{3\xspace}
\newcommand{\rHttpCves}{41\xspace}
\newcommand{\rIvCves}{7\xspace}
\newcommand{\rIvExploit}{142\xspace}
\newcommand{\rIvExploitPay}{133\xspace}
\newcommand{\rIvLegit}{5{,}587\xspace}

\newcommand{\rJfDarpa}{7\xspace}

\newcommand{\rJfFieldsAccept}{219\xspace}

\newcommand{\rJfHardAccept}{130\xspace}
\newcommand{\rJfHardNeg}{2{,}048\xspace}
\newcommand{\rJfNeg}{4{,}096\xspace}

\newcommand{\rJfOperandsPct}{45\xspace}
\newcommand{\rLabCaptured}{10\xspace}

\newcommand{\rLabHostMedian}{3{,}127\xspace}
\newcommand{\rLabReqTotal}{147\xspace}
\newcommand{\rLabWin}{2\xspace}

\newcommand{\rLgCaps}{9\xspace}

\newcommand{\rLgRequests}{529\xspace}

\newcommand{\rLibFamilies}{45\xspace}
\newcommand{\rLibFpEpmm}{two\xspace}
\newcommand{\rLibFpIvanti}{four\xspace}

\newcommand{\rLibGridFp}{5\xspace}
\newcommand{\rLibGridPairs}{1{,}964\xspace}
\newcommand{\rLibGridPossible}{1{,}980\xspace}

\newcommand{\rLibPairsSkipped}{16\xspace}
\newcommand{\rLibQuiet}{40\xspace}

\newcommand{\rLlmAlertHits}{12\xspace}
\newcommand{\rLlmAlertOf}{51\xspace}

\newcommand{\rLlmEncBenign}{0\xspace}
\newcommand{\rLlmGptBenign}{12\xspace}
\newcommand{\rLlmGptEv}{7\xspace}
\newcommand{\rLlmGptEvWrong}{3\xspace}
\newcommand{\rLlmGptFound}{31\xspace}
\newcommand{\rLlmGptWrong}{31\xspace}
\newcommand{\rLlmOpusBenign}{12\xspace}
\newcommand{\rLlmOpusEv}{9\xspace}
\newcommand{\rLlmOpusEvWrong}{2\xspace}
\newcommand{\rLlmOpusFound}{32\xspace}
\newcommand{\rLlmOpusWrong}{37\xspace}
\newcommand{\rLocCves}{40\xspace}
\newcommand{\rLocFieldEventsMax}{600\xspace}
\newcommand{\rLocFieldEventsMed}{10\xspace}

\newcommand{\rLocFieldProcsOverOne}{15\xspace}

\newcommand{\rLocWitEventsMed}{2\xspace}

\newcommand{\rLtDriven}{8\xspace}

\newcommand{\rLtMinutes}{5\xspace}

\newcommand{\rLtRequests}{23{,}512\xspace}
\newcommand{\rLtSilentApps}{3\xspace}

\newcommand{\rMcMaxP}{0.008\xspace}

\newcommand{\rNoRuleCves}{10\xspace}

\newcommand{\rNoRuleWeb}{7\xspace}

\newcommand{\rNuFoundReached}{5\xspace}

\newcommand{\rNuReached}{6\xspace}

\newcommand{\rPaIdent}{363\xspace}
\newcommand{\rPaKept}{64\xspace}

\newcommand{\rPayDetLab}{7\xspace}

\newcommand{\rPiIdentifiers}{363\xspace}

\newcommand{\rPostFreeze}{10\xspace}
\newcommand{\rPostFreezeDet}{9\xspace}

\newcommand{\rPrExchanges}{143\xspace}
\newcommand{\rPrPayCves}{7\xspace}
\newcommand{\rPrPayExchanges}{134\xspace}
\newcommand{\rPsAvail}{30\xspace}
\newcommand{\rPsEncAny}{28\xspace}

\newcommand{\rPsRest}{15\xspace}
\newcommand{\rPsRuleFired}{30\xspace}

\newcommand{\rRcCvesWeb}{28\xspace}

\newcommand{\rRcRequestLine}{28\xspace}

\newcommand{\rRcSameEndpoint}{20\xspace}
\newcommand{\rRcWebRules}{31\xspace}
\newcommand{\rRgAllAlertsLt}{0\xspace}

\newcommand{\rRgAllAlertsOther}{0\xspace}

\newcommand{\rRgAllLeads}{9{,}113\xspace}

\newcommand{\rRgCfXaWrong}{254\xspace}

\newcommand{\rRgEndLeads}{6{,}321\xspace}

\newcommand{\rRgEndLeadsLt}{6{,}219\xspace}

\newcommand{\rRgParAdd}{958\xspace}

\newcommand{\rRgRootAdd}{1{,}834\xspace}

\newcommand{\rRgSust}{23{,}512\xspace}
\newcommand{\rRgUnits}{27{,}081\xspace}
\newcommand{\rRlCrawl}{564\xspace}
\newcommand{\rRlCves}{3\xspace}

\newcommand{\rRlLegit}{172\xspace}
\newcommand{\rRlRuleFlag}{27\xspace}

\newcommand{\rRqTwoAnchors}{136\xspace}
\newcommand{\rRqTwoCves}{43\xspace}

\newcommand{\rRqTwoDescSup}{5\xspace}

\newcommand{\rRqTwoSupList}{\CodeIn{actuator}, \CodeIn{remote\_agent.php}, \CodeIn{template}, \CodeIn{cli}, \CodeIn{geoserver}\xspace}
\newcommand{\rRqTwoZero}{38\xspace}

\newcommand{\rRuleOther}{5\xspace}

\newcommand{\rSbHosts}{5\xspace}

\newcommand{\rSbRuleAlerts}{489\xspace}

\newcommand{\rSbRuleEvents}{224{,}046\xspace}
\newcommand{\rSbRuleFired}{19\xspace}

\newcommand{\rSbWindows}{830\xspace}
\newcommand{\rSensDeltaMax}{60\xspace}
\newcommand{\rSensDeltaMin}{5\xspace}
\newcommand{\rSensEpsMax}{10\xspace}
\newcommand{\rSensEpsMin}{0\xspace}
\newcommand{\rSensHMax}{4\xspace}
\newcommand{\rSensHMin}{2\xspace}
\newcommand{\rSensKMax}{5\xspace}
\newcommand{\rSensKMin}{1\xspace}

\newcommand{\rStCvesWithRules}{41\xspace}
\newcommand{\rStFound}{35\xspace}

\newcommand{\rStPayFound}{5\xspace}
\newcommand{\rStRules}{114\xspace}
\newcommand{\rStThird}{25\xspace}
\newcommand{\rStWrong}{3\xspace}

\newcommand{\rSubsetCves}{30\xspace}

\newcommand{\rTierAnyCross}{5\xspace}
\newcommand{\rTierAnyDet}{43\xspace}

\newcommand{\rTierAnyThird}{31\xspace}
\newcommand{\rTierEvCross}{0\xspace}
\newcommand{\rTierEvDet}{12\xspace}

\newcommand{\rTierHostCves}{4\xspace}
\newcommand{\rTierLabAny}{12\xspace}
\newcommand{\rTierLabEv}{9\xspace}

\newcommand{\rTierPayCves}{8\xspace}
\newcommand{\rTierThirdEv}{3\xspace}
\newcommand{\rTpEvPerSec}{222\xspace}
\newcommand{\rTpHostDayMin}{2.65\xspace}
\newcommand{\rTpHostsPerCore}{543\xspace}

\newcommand{\rTpMembers}{443\xspace}
\newcommand{\rTpRssMb}{57\xspace}
\newcommand{\rTpWinEvents}{368\xspace}
\newcommand{\rTpWinHourEvPerSec}{210\xspace}
\newcommand{\rTpWinHourEvents}{1{,}437\xspace}
\newcommand{\rTpWinHourSec}{6.8\xspace}
\newcommand{\rTpWinSec}{1.66\xspace}

\newcommand{\rXaCqWrong}{293\xspace}

\newcommand{\rZpLab}{9\xspace}

\newcommand{\rZpRequests}{1{,}083\xspace}
\newcommand{\rZpSimilar}{8\xspace}

\newcommand{\rEcoRulePocFirst}{33\xspace}
\newcommand{\rEcoRuleDated}{37\xspace}
\newcommand{\rEcoCves}{7{,}031\xspace}
\newcommand{\rEcoKevMapped}{651\xspace}
\newcommand{\rEcoKevParses}{729\xspace}
\newcommand{\rEcoKevPoc}{942\xspace}
\newcommand{\rEcoKevPython}{771\xspace}
\newcommand{\rEcoKevWindow}{1{,}186\xspace}
\newcommand{\rEcoLagMedian}{12\xspace}
\newcommand{\rEcoLeCritShare}{67.81\%\xspace}
\newcommand{\rEcoLeKevPython}{4{,}554\xspace}
\newcommand{\rEcoLeShare}{60.02\%\xspace}
\newcommand{\rEcoMapShare}{88.4\%\xspace}
\newcommand{\rEcoPocs}{14{,}933\xspace}
\newcommand{\rEcoPythonPocs}{5{,}776\xspace}
\newcommand{\rEcoPythonShare}{38.7\%\xspace}
\newcommand{\rEcoRuleLagMedian}{22\xspace}

\newcommand{\rTableMatchBody}{%
\multirow{3}{*}{\parbox{1.7cm}{\centering E3\\ Cadets}} & \poirot & 501 & 246 & 11 & 266 & 0.749 & 0.671 & 0.979 & 0.796 \\
 & \provgs & 512 & 187 & 0 & 325 & 0.817 & 0.732 & 1.000 & 0.846 \\
 & \cellcolor{g2}\Sys & \cellcolor{g2}512 & \cellcolor{g2}15 & \cellcolor{g2}0 & \cellcolor{g2}497 & \cellcolor{g2}0.985 & \cellcolor{g2}0.972 & \cellcolor{g2}1.000 & \cellcolor{g2}0.986 \\
\midrule
\multirow{3}{*}{\parbox{1.7cm}{\centering E3\\ Theia}} & \poirot & 432 & 227 & 80 & 285 & 0.700 & 0.656 & 0.844 & 0.738 \\
 & \provgs & 512 & 162 & 0 & 350 & 0.842 & 0.760 & 1.000 & 0.863 \\
 & \cellcolor{g2}\Sys & \cellcolor{g2}512 & \cellcolor{g2}28 & \cellcolor{g2}0 & \cellcolor{g2}484 & \cellcolor{g2}0.973 & \cellcolor{g2}0.948 & \cellcolor{g2}1.000 & \cellcolor{g2}0.973 \\
\midrule
\multirow{3}{*}{\parbox{1.7cm}{\centering E3\\ Trace}} & \poirot & 502 & 52 & 10 & 460 & 0.939 & 0.906 & 0.980 & 0.942 \\
 & \provgs & 511 & 172 & 1 & 340 & 0.831 & 0.748 & 0.998 & 0.855 \\
 & \cellcolor{g2}\Sys & \cellcolor{g2}512 & \cellcolor{g2}5 & \cellcolor{g2}0 & \cellcolor{g2}507 & \cellcolor{g2}0.995 & \cellcolor{g2}0.990 & \cellcolor{g2}1.000 & \cellcolor{g2}0.995 \\
\midrule
\multirow{3}{*}{\parbox{1.7cm}{\centering E3\\ FiveDirections}} & \poirot & 354 & 34 & 158 & 478 & 0.812 & 0.912 & 0.691 & 0.787 \\
 & \provgs & 512 & 210 & 0 & 302 & 0.795 & 0.709 & 1.000 & 0.830 \\
 & \cellcolor{g2}\Sys & \cellcolor{g2}512 & \cellcolor{g2}6 & \cellcolor{g2}0 & \cellcolor{g2}506 & \cellcolor{g2}0.994 & \cellcolor{g2}0.988 & \cellcolor{g2}1.000 & \cellcolor{g2}0.994 \\
\midrule
\multirow{2}{*}{\parbox{1.7cm}{\centering E5\\ Cadets}} & \poirot & 262 & 116 & 250 & 396 & 0.643 & 0.693 & 0.512 & 0.589 \\
 & \cellcolor{g2}\Sys & \cellcolor{g2}512 & \cellcolor{g2}35 & \cellcolor{g2}0 & \cellcolor{g2}477 & \cellcolor{g2}0.966 & \cellcolor{g2}0.936 & \cellcolor{g2}1.000 & \cellcolor{g2}0.967 \\
\midrule
\multirow{2}{*}{\parbox{1.7cm}{\centering E5\\ Theia}} & \poirot & 254 & 119 & 258 & 393 & 0.632 & 0.681 & 0.496 & 0.574 \\
 & \cellcolor{g2}\Sys & \cellcolor{g2}512 & \cellcolor{g2}33 & \cellcolor{g2}0 & \cellcolor{g2}479 & \cellcolor{g2}0.968 & \cellcolor{g2}0.939 & \cellcolor{g2}1.000 & \cellcolor{g2}0.969 \\
\midrule
\multirow{2}{*}{\parbox{1.7cm}{\centering E5\\ ClearScope}} & \poirot & 505 & 202 & 7 & 310 & 0.796 & 0.714 & 0.986 & 0.829 \\
 & \cellcolor{g2}\Sys & \cellcolor{g2}512 & \cellcolor{g2}7 & \cellcolor{g2}0 & \cellcolor{g2}505 & \cellcolor{g2}0.993 & \cellcolor{g2}0.987 & \cellcolor{g2}1.000 & \cellcolor{g2}0.993 \\
\midrule
\multirow{2}{*}{\parbox{1.7cm}{\centering ATLASv2}} & \poirot & 413 & 164 & 99 & 348 & 0.743 & 0.716 & 0.807 & 0.758 \\
 & \cellcolor{g2}\Sys & \cellcolor{g2}512 & \cellcolor{g2}58 & \cellcolor{g2}0 & \cellcolor{g2}454 & \cellcolor{g2}0.943 & \cellcolor{g2}0.898 & \cellcolor{g2}1.000 & \cellcolor{g2}0.946 \\}

\newcommand{\rPcCqLab}{9\xspace}
\newcommand{\rPcCqThird}{25\xspace}

\newcommand{\rPcSplFound}{32\xspace}
\newcommand{\rPcSplLab}{2\xspace}
\newcommand{\rPcSplThird}{30\xspace}
\newcommand{\rPcStLab}{10\xspace}

\renewcommand\footnotetextcopyrightpermission[1]{}

\begin{document}

\title{Set an Exploit to Catch an Exploit: Compiling Public Proofs of Concept into Threat Hunting Rules}

\author{Muhammad Shoaib}
\affiliation{%
  \institution{University of Virginia}
  \city{Charlottesville}
  \country{USA}}
\author{Alexander Suh}
\affiliation{%
  \institution{University of Virginia}
  \city{Charlottesville}
  \country{USA}}
\author{\texorpdfstring{\mbox{Joshua Andrew Tomiak}}{Joshua Andrew Tomiak}}
\affiliation{%
  \institution{University of Virginia}
  \city{Charlottesville}
  \country{USA}}
\author{Wajih Ul Hassan}
\affiliation{%
  \institution{University of Virginia}
  \city{Charlottesville}
  \country{USA}}
\renewcommand{\shortauthors}{Shoaib et al.}
\settopmatter{authorsperrow=4}

\begin{abstract}
Public exploit code for a newly disclosed vulnerability usually appears before any published rule
that detects its exploitation. Of the 45 \cves we evaluate, 37 ever received a
Sigma, Elastic or Splunk rule; in 33 the exploit came first, by a median of
22 days. Analysts still write every rule by hand. We present \Sys, which compiles a
public Python proof-of-concept exploit (\poc) into \cve-specific hunting rules over a web service's existing logs, without running it. \Sys rests on the \poc Behavior Alignment Principle: a \poc
performs the exploit, so its code already provides a rule's HTTP requests, payloads and host operations. That is hard to act on: a \poc is written against libraries and runs on the attacker's
machine, while the target's web-server, proxy and host logs record only what arrived and what it did.
\Sys closes the gap with a map from library calls to the operations they cause, built once per library, keeping only the operations visible on the target. For 12 of the 45, \sys's rule states the exploit's own payload fragment or host
operation, which is evidence of the exploit itself, and not of traffic to its endpoint. On matched controls it separates that exploit from the same request with
a harmless value, and it raises no alerts across 27{,}081 legitimate and crawl requests. We never compile the exploit that produced the logs. Most are Splunk's
public attack data, recorded by a third party. Without reading a single log, \sys's rules find 43 of the 45. Splunk's log-tested rules cover 30 of the 45, and \sys finds 28 of those, plus all 10 \cves without an ESCU rule. Where Splunk's rules alert on 20 wrong \cve pairs, \sys's alert on only 2. Hunting can start as soon as a \poc is public, and rules can raise an alert when
it provides a payload or host operation.
\end{abstract}



\maketitle

\section{Introduction}
\label{s:intro}

Public exploit code for a newly disclosed vulnerability usually appears before any rule that
detects its use. For high- or critical-severity \cves on CISA's list of known exploited
vulnerabilities~\cite{highriskcvecisa}, a public proof-of-concept exploit (\poc) appeared a median of
\rEcoLagMedian days after disclosure. Of the \rCorpusTotal evaluated \cves, \rEcoRuleDated have a published Sigma, Elastic or Splunk rule for access or host logs. For \rEcoRulePocFirst of them, a Python \poc was
public a median of \rEcoRuleLagMedian days before the first rule (\secref{sec:cve:ecosystem}). After Heartbleed,
patching of popular HTTPS sites plateaued after about two weeks, while exploit attempts began within a
day~\cite{durumeric2014heartbleed}.

Most of the evaluated \cves (\rHttpCves of \rCorpusTotal) affect web applications or web-facing appliances such as VPN gateways and file-transfer servers. Attackers exploit them with HTTP requests, which appear in web-server and proxy access logs and in host logs. Today, detection engineers write rules over these logs by hand, one
\cve at a time, in Sigma~\cite{sigmasignatures} or in the search language of a security information and
event management (SIEM) system~\cite{splunk}. Provenance query systems~\cite{poirot2019,altinisik2023provg}
match an analyst's hand-written query graph. Query extractors~\cite{satvat2021extractor,gao2021enabling}
build queries from attack reports, but the public description of a new \cve mentions only \rRqTwoDescSup of the \rRqTwoAnchors operand values in our rules (\secref{eval:rq2}). Systems that turn exploit
code into signatures~\cite{wang2013metasymploit,kobayashi2023idssig} produce payload patterns for network
intrusion detectors and do not cover access logs or host activity.

\PPnew{The \poc Behavior Alignment Principle.} \emph{A \poc is a program that performs the exploit,
so a rule for its \cve must match the \poc's requests, payloads and host operations on the target, and aligning these with a deployment's logs finds attempts to exploit that \cve.} To trigger the vulnerability, a \poc's author has to spell out the URL path, the payload parameter, the uploaded file and the server-side command. Independent authors' \pocs for the same \cve target the same endpoints (\secref{eval:rq4}).

A \poc usually runs on the attacker's machine and calls libraries such as \CodeIn{requests}, while defenders see the attack in the target's logs. On the target, an HTTP call from the \poc produces an access-log entry and a file read by the server process. Many values in a rule's predicates are computed at run time, and the \poc also sends strings unrelated to the exploit. A compiler therefore has to resolve these values and tell the exploit's strings apart from unrelated ones.
A widely used static analyzer~\cite{codeql,avgustinov2016ql} finds every hard-coded string in a \poc. When we turned each string into a rule (\sysCQ), the rules alerted on \rCqLtContent legitimate requests (\secref{eval:parts}), because strings such as a JSON content type or a browser's user agent also appear in ordinary traffic. \Sys's payload and host rules raise no alert on these requests.

\Sys compiles a public Python \poc into \cve-specific hunting rules over the access and host logs already collected by web operators. Without running the \poc, \sys looks up the operations of each library call in a map, resolves the values passed by the \poc and keeps the operations that occur on the target. It translates these operations into the target's log records (\secref{s:design}).
A \emph{request rule} matches the HTTP request to the vulnerable endpoint, in the same way as an analyst's web rule. A \emph{payload rule} also matches the content of that request, and a \emph{host rule} matches the exploit's operations on the server. Each rule is a conjunction of field predicates that an existing SIEM can run as a correlation search. A rule flags failed exploit attempts as well as successful ones. Because \sys needs only the \poc, defenders can start hunting as soon as a Python \poc for a \cve is public.

We evaluate \Sys on \rCorpusTotal \cves, \rCorpusSplunk of them on Splunk's public attack data, recorded by a third party with exploit code unknown to us. Payload and host rules detect the exploits of \rTierEvDet \cves with evidence and raise no alert on \rRgUnits requests from legitimate use and crawling. With request rules, which add leads for an analyst, \sys's rules detect the exploits of \rTierAnyDet \cves. Splunk tested its rules on these logs, and we compiled ours without seeing any log. Still, \sys's rules alert on another \cve's logs \rCcFold times less often than Splunk's (\rCcEncFp versus \rCcRuleFp \cve pairs) and detect all \rNoRuleCves \cves that have no Splunk rule. With the static analyzer or our simpler extractor in place of \sys's extraction and target-side translation, the rules detect \rBcCqOwn and \rBcStOwn \cves, respectively.

We make four contributions:
\begin{itemize}[leftmargin=*]
\item The \poc Behavior Alignment Principle, and a measurement of how long public \pocs precede published detection rules (\secref{sec:cve:ecosystem}).
\item \Sys, a compiler from Python \pocs to request, payload and host rules (\secref{s:design}).
\item An evaluation on \rCorpusTotal \cves against Splunk's deployed rules and two simpler versions of \sys (\secref{s:eval}).
\item An artifact with the compiler, the matcher and the scripts behind our evaluation (Appendix~\ref{app:openscience}).
\end{itemize}

\section{Background}
\label{s:prelim}

\subsection{Event Graphs}
Enterprises already log host and web activity. Host collectors such as \CodeIn{auditd}~\cite{auditd} and Sysmon record process, file and network operations, network monitors such as Zeek~\cite{zeek} record connections, and web servers write each request to an access log. We turn these logs into a directed multigraph \evtg, the \ldeg. Its vertices are processes, files, network endpoints and requested resources, labelled with an executable name, a path or an address. Each edge is one timestamped operation with an \emph{event class} from
$\Sigma_\sigma=\{\textsc{create},\allowbreak\textsc{read},\allowbreak\textsc{write},\allowbreak\textsc{connect},\allowbreak\textsc{access},\allowbreak\textsc{recv}\}$.
A request becomes an \textsc{access} edge from the client to the serving process and an edge from that process to the requested resource, a \textsc{read} for methods such as GET and a \textsc{write} for the others. We label the resource vertex with the raw request path, the percent-decoded path and the request's operands. If the log also has the query, body or headers, we add a \textsc{recv} edge from the serving process to that content. We attach request and host records to the same serving-process vertex by service name, and we treat workers that share a listening socket as one process.

\subsection{Rules as Templates}
A template \qg is a small graph with the vertex types and event classes of \evtg. Each vertex label is a concrete operand, a wildcard or a disjunction of patterns. An operand is a value in the \poc's code, such as a URL path, file name or address. We call a vertex with a concrete operand an \emph{anchor}. We compile one \cve's \poc into a \emph{family} of templates, each a request, payload or host \emph{rule} (\secref{s:intro}). A rule
\emph{aligns} to \evtg when an injective mapping from its vertices to those of \evtg preserves type and
label and maps every template edge to an edge of \evtg or to a path through at most $k$ intermediate
process creations (\secref{sec:refine}). Alignment is thus subgraph containment with a bounded path relaxation, a problem also solved by query-based provenance detectors~\cite{poirot2019,altinisik2023provg}. A
\textsc{remote} \poc runs on the attacker's machine and connects to the target over the network, and a
\textsc{local} one runs on the vulnerable host (\secref{s:design:transfer}).

\subsection{Threat Model}
\label{s:threat}
The attacker exploits a disclosed \cve that has a public \poc, against a host or service monitored by the defender. The attacker may work from any network position and may use that \poc or another implementation of the same exploit. The defender collects standard host, network and web-server logs and installs nothing specific to the vulnerable application. Following prior provenance work~\cite{king2003,inam2022sok}, we trust the logging pipeline and assume that records are collected before the attacker can alter them. \Sys looks only for \cves with compiled \pocs and alerts when one of their rules aligns to the logs. It needs a public Python \poc for each target \cve, and it does not compile rules for raw-socket protocols outside the API-to-operation map $\Pi$ (\secref{s:design:apitocalls}).

\section{The \texorpdfstring{\poc}{PoC} Ecosystem}
\label{sec:cve:ecosystem}

We matched \cve metadata from CVEdetails~\cite{cvedetails} to the \pocs in four public sources~\cite{exploitdb,RhinoSecurityLabs,Kernelhub,pocingithub} and obtained \rEcoCves \cves and \rEcoPocs \pocs. We exclude hardware, mobile and niche-OS vulnerabilities. All counts come from a snapshot taken on 30 September 2025, when every evaluated \cve and its \poc were public. Among the high- or critical-severity (CVSS 7.0 or above) \cves in CISA's Known Exploited Vulnerabilities (KEV) catalog~\cite{highriskcvecisa} that have a public \poc in our sources, the \poc appeared a median of \rEcoLagMedian days after disclosure. Of the \rCorpusTotal evaluated \cves (\secref{eval:setup}), \rEcoRuleDated have a published Sigma, Elastic or Splunk rule for access or host logs, and for \rEcoRulePocFirst of them a Python \poc was public a median of \rEcoRuleLagMedian days before the first rule.

We wrote \sys's front end for Python because it is the most common language in the corpus, used by \rEcoPythonPocs of the \rEcoPocs \pocs (\rEcoPythonShare). Collectors log process, file or network events from \rEcoLeShare of all \pocs and \rEcoLeCritShare of those for \cves with a CVSS score above 9.0. Among the Python \pocs for KEV-listed \cves, \rEcoLeKevPython produce such events. $\Pi$ maps every log-relevant call of the Python \poc for \rEcoKevMapped of the \rEcoKevWindow KEV-listed \cves disclosed since 2019 (Appendix~\ref{app:rq4detail}).

\section{Design: \texorpdfstring{\Sys}{Encore}}
\label{s:design}

\begin{figure*}[!t]
    \centering
    \includegraphics[width=\textwidth]{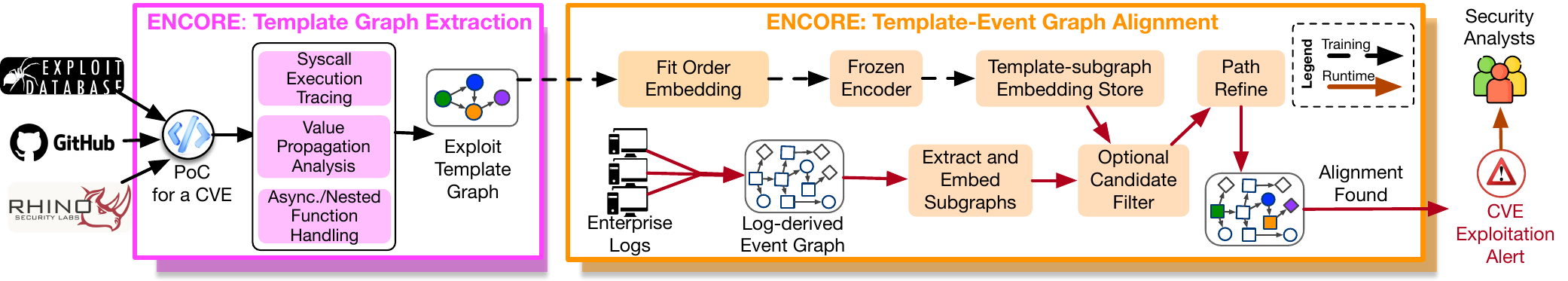}
        \caption{Workflow of \Sys. We trace each library's calls once, offline, and record their operations in the API-to-operation map $\Pi$ (\secref{s:design:apitocalls}). \Sys aligns templates with the \ldeg (\evtg) by path refinement. The order-embedding boxes show an optional learned filter that prunes candidates on large provenance graphs (Appendix~\ref{app:screen}).}
  \label{fig:arch} \vspace{-2ex}
\end{figure*}

A \poc calls library functions, but the target's logs show process, file and network operations and HTTP requests. We connect the two with the API-to-operation map $\Pi$ (\secref{s:design:apitocalls}), which lists the events produced by each library call. \Sys analyzes a \poc statically and needs no live target (\secref{s:design:pipeline}). It looks up each call of the \poc in $\Pi$ and builds a graph with the same vertex and edge types as the event graph~(Figure~\ref{fig:arch}).

\subsection{API-to-Operation Map}
\label{s:design:apitocalls}
$\Pi$ maps a normalized API identifier to an ordered list of \emph{operation records}. Each record lists an event class $\sigma$ in $\Sigma_\sigma$ (\secref{s:prelim}) and the acting process's subject reference $\mathit{ref}$, either \textsc{Caller} or $\textsc{Created}(j)$. It also lists a role and field map for each subject and object, linking each field to an \emph{argument selector}, a path inside an argument value. When we trace syscalls offline, a typing map $T$ assigns each syscall a vertex type and a direction, and a lifting $L$ assigns it an event class.

To measure part of $\Pi$, we run each library call under a host collector and record the resulting syscalls (Appendix~\ref{app:pi}). This tracing of library calls is \sys's only dynamic analysis. A trace shows a call's \CodeIn{sendto}, but nothing in it maps \CodeIn{url.path} to the request path. We therefore write the rest of $\Pi$ by hand: the field maps, the library transformations from arguments to bytes, and one selector-to-field map per log format. $\Pi$ maps \rPiIdentifiers identifiers and covers \rEcoMapShare of the normalized library and builtin calls in the \poc corpus. We spent roughly 9 person-days building it, about half of them on field maps, and we need about one more person-day for each new (OS, collector, runtime) tuple.

We froze $\Pi$ on 1 March 2024. Of the evaluated \cves, \rPostFreeze were disclosed after that date, and the rules find \rPostFreezeDet of them. We fixed the path levels, the common-directory rule, the fragment test, the usage-text fallback, the alert predicate's weights and the alignment tolerances before selecting the \cve corpus, and we never added or retuned a compiler heuristic for a \cve in the corpus (Appendix~\ref{app:freeze}).

\subsection{Target-Side Translation}
\label{s:design:transfer}
The target's logs contain none of the \poc's syscalls on the attacker's machine. \Sys therefore builds vertices only from the \poc's \emph{operands} and from its \emph{operations} on the target machine. \Sys locates each operand in a call's arguments with $\Pi$'s selector and resolves it to a literal in the \poc by value propagation. It then applies the library transformation in $\Pi$ to the argument and checks the result against the corresponding field in the collector profile.

Authors seldom mark a \poc as \textsc{local} or \textsc{remote}, so \sys compiles every \poc in both modes and deploys all the rules. In \textsc{remote} mode, \sys uses $R^*$ to translate each call to the target into the target's log record, such as an access-log entry for an HTTP request. The \emph{method map} turns \CodeIn{GET}, \CodeIn{HEAD} and \CodeIn{OPTIONS} requests into \textsc{read} edges and other requests into \textsc{write} edges. \Sys discards calls to other hosts, which affect only the attacker's machine. In \textsc{local} mode, it keeps the \poc's file, process and socket operations. \Sys checks a rule only where the target's logs include all its event classes and requires a match for every edge (Appendix~\ref{app:observe}). Table~\ref{tab:collectors} lists \sys's collector profiles.

\begin{table}[t]
\tablestyle\centering
\caption{The three collector profiles, with the events recorded by each and whether request rules apply.}
\label{tab:collectors}
\begin{tabular}{@{}>{\raggedright\arraybackslash}p{0.15\columnwidth}
                >{\raggedright\arraybackslash}p{0.34\columnwidth}
                >{\raggedright\arraybackslash}p{0.21\columnwidth}
                >{\raggedright\arraybackslash}p{0.19\columnwidth}@{}}
\toprule
Profile & Collectors & Records & Request rules \\
\midrule
Host & Sysmon, auditd, Process Monitor, a syscall sidecar & process, file and connection events & no \\
\addlinespace[2pt]
Network & Zeek, Suricata, firewall logs & connections and parsed requests & where requests are logged \\
\addlinespace[2pt]
Application & web access, file-transfer and audit logs & requests & yes, with bodies where logged \\
\bottomrule
\end{tabular} \vspace{-2ex}
\end{table}

\begin{figure*}[!t]
    \centering
    \includegraphics[width=\textwidth]{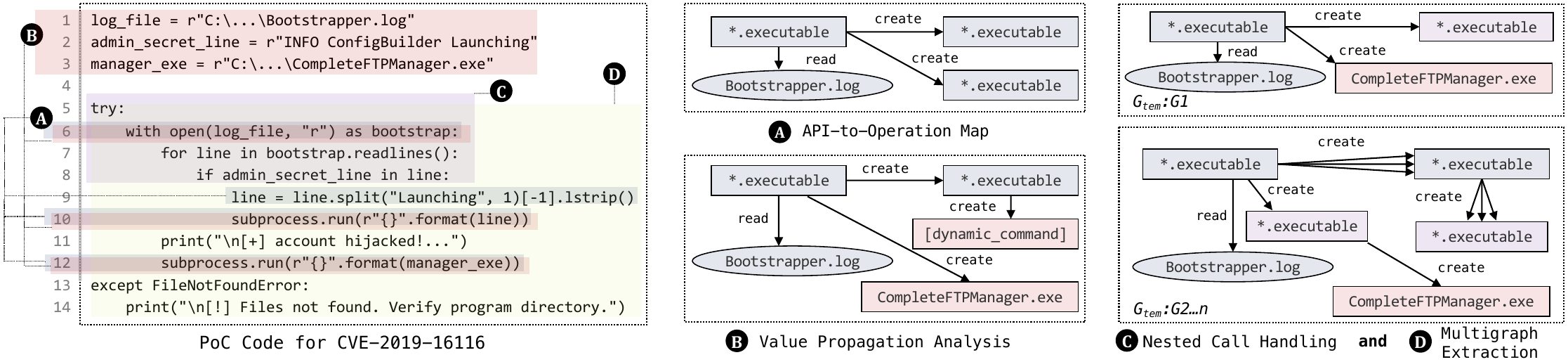}
    \caption{Template (\qg) generation for CVE-2019-16116: (A) the API-to-operation map
    $\Pi$, (B) value propagation, (C) nested-call handling, (D) multigraph extraction.}
    \label{fig:qg:capabilities}
    \vspace{-2ex}
\end{figure*}

\subsection{Extracting a Template}
\label{s:design:pipeline}
\Sys parses a \poc into a Python AST and follows import aliases and method chains to give each call site a stable API identifier, so that \CodeIn{requests.get} and \CodeIn{requests.Session().get} map to the same entry in $\Pi$. A sanitizer then prunes the \poc to the statements on which a call in $\Pi$ depends. The \emph{call-chain sanitizer} also keeps calls to functions defined in the \poc that reach such a call. We compile every \poc with both sanitizers. It then propagates values through assignments and simple string expressions to fill in the record fields of each mapped call. The compiler uses a field's resolved literal as the vertex label, or a wildcard otherwise. Two wildcards share a symbolic variable only if each spans a whole field and both fields have the same selector path and library transformation (Appendix~\ref{app:designdetail}). The compiler inlines the functions defined in the \poc, including coroutines, at their call sites and binds arguments to parameters. It stops at depth 3 and does not inline recursive calls. We also report every call missing from $\Pi$ or hidden from the compiler (Appendix~\ref{app:algo}).

The compiler turns the \poc of a \cve $c$ into a finite family $\mathcal{G}_c$ of canonical template graphs, with one graph for each distinct execution path through the \poc (up to 256) and one for each \emph{variant} of a path. A variant is a copy of a path with some requests removed or with an extra edge for a request's content. \Sys alerts for $c$ when a rule $g\in\mathcal{G}_c$ matches a candidate by refinement and meets the alert predicate in \secref{sec:refine}. We apply the predicate to one rule at a time.
Figure~\ref{fig:qg:capabilities} shows template extraction for the \poc of CVE-2019-16116, a local-mode exploit that sends nothing over the network. \Sys recovers the concrete labels \CodeIn{Bootstrapper.log} and \CodeIn{CompleteFTPManager.exe} by propagating values. The command, built at run time from the contents of a file, becomes a wildcard that depends on that file read.

\subsection{Refinement and Scoring}
\label{sec:refine}
\Sys checks each rule against \emph{candidates}\label{sec:align:ml}, the subgraphs of \evtg within $h$ hops of a process vertex (Appendix~\ref{s:align}). Refining a rule on a candidate means finding an injective vertex mapping $\mu_V: V' \to V$ and an injective terminal-edge mapping $\mu_E: E' \to E$ from the template's vertices $V'$ and edges $E'$ to the candidate's $V$ and $E$ (Appendix~\ref{app:aligndetail}). Each template edge $e=(u,v,\sigma)$ maps to a directed path $P_e$ that starts at $\mu_V(u)$, ends with an event $\mu_E(e)$ of class $\sigma$ at $\mu_V(v)$, and passes through at most $k$ intermediate process creations. On a request edge, the path may also start at another process of the same service that shares $\mu_V(u)$'s listening socket. On such edges, \textsc{read} and \textsc{write} count as one class, because an exploit may send the same request with another method. Matched edges must follow the compiler's emission order, up to a clock skew of $\epsilon$, and span at most $\Delta$ (Appendix~\ref{app:timing}). \Sys never abstracts or merges process, file or socket vertices. \Sys refines every candidate in a \cve's logs, and on large provenance graphs an optional learned filter prunes candidates before refinement (Appendix~\ref{app:screen}).

\Sys alerts on a rule when such a mapping exists and the rule contains at least three vertices, or two vertices if one of them has a concrete label. A vertex with a concrete label has weight $0.33$, and a vertex with a wildcard, a placeholder or the \poc's executable has weight $0.20$. The mapped vertices must have a total weight of at least $0.47$. We set $k=3$, $h=3$, $\Delta=15$ minutes and $\epsilon=2$ seconds. When we vary one tolerance at a time ($k$ from \rSensKMin to \rSensKMax, $h$ from \rSensHMin to \rSensHMax, $\Delta$ from \rSensDeltaMin to \rSensDeltaMax minutes and $\epsilon$ from \rSensEpsMin to \rSensEpsMax seconds), the rules find the same \rTierAnyDet \cves, \rTierEvDet of them with evidence, and alert on the same \rTierAnyCross wrong-\cve pairs (\secref{eval:rq1}).

\subsubsection{Rule Structure}
A request rule matches the requested resource at up to three \emph{levels}: the requested \emph{endpoint}, its parent directory and the application's context root. Rules at the higher levels also match exploits that request other resources of the same component. Under the \emph{common-directory rule}, we drop any rule whose path is a single component from SecLists' list of common web directory names (\CodeIn{common.txt})~\cite{seclists}. When a \poc's request contains a literal fragment, such as a command or an expression, that passes the \emph{fragment test} (Appendix~\ref{app:designdetail}), the compiler adds \emph{payload rules} to the family. Each payload rule is a request rule with an extra \textsc{recv} edge for request content that contains the fragment. If a collector logs each exchange as one record, this content must come from the matched request.

\section{Evaluation}
\label{s:eval}

\PP{Setup.}\label{eval:setup} We evaluate on \rCorpusTotal \cves disclosed from 2019 to 2025. The exploit in each \cve's logs came from code other than the compiled \poc. For \rCorpusSplunk of these \cves, the logs are Splunk's public \CodeIn{attack\_data}~\cite{splunkattackdata}, recorded by a third party. We never saw the code that produced them and use them unmodified. We recorded the logs of the other \rCorpusLab: \rLabCaptured web \cves on vulhub builds~\cite{vulhub} and \rLabWin Windows \cves. We deployed each vulnerable build and exploited it with another author's public \poc, from a repository that shares no fork or upstream with the compiled one (Appendix~\ref{app:setupdetail}). For each \cve, we compile one public Python \poc that sends the triggering request. The exception is ColdFusion's CVE-2023-29298, whose closest public code only logs in as administrator. Without this \cve, the rules find \rExSubDet of the remaining \rExSubTotal. On the \rLabCaptured web \cves, which we call \emph{instrumented}, a reverse proxy in the role of a WAF or API gateway records each HTTP exchange with its request body, and a sidecar in the role of an endpoint agent records the server's process, file and socket syscalls. On the \rLabWin Windows \cves, Process Monitor~\cite{procmon} and Sysmon~\cite{sysmon} record these operations.

\begin{table}[t]
\tablestyle\centering
\caption{All systems on the same logs. \sysCQ uses CodeQL in place of \sys's extraction, and \sysEX uses our extractor (\secref{eval:parts}). Wrong-\cve pairs (of \rLibGridPairs): one \cve's rule alerts on another's logs. n/a: not scored (\secref{eval:rq1}).}
\label{tab:compare}
\resizebox{\columnwidth}{!}{%
\begin{tabular}{@{}lrrrrr@{}}
\toprule
& \multicolumn{3}{c}{\cves found} & & \\
\cmidrule(lr){2-4}
& of \rCorpusTotal & third-party & our runs & Evidence & \smash{\begin{tabular}[b]{@{}r@{}}Wrong-\cve\\pairs\end{tabular}} \\
\midrule
Splunk's \cve rules & \rPcSplFound & \rPcSplThird & \rPcSplLab & n/a & n/a \\
\sysCQ & \rCqDet & \rPcCqThird & \rPcCqLab & \rCqContentEv & \rCqWrong \\
\quad with filters & \rCqDet & \rPcCqThird & \rPcCqLab & \rCqRungEv & \rCqRungWrongAll \\
\sysEX & \rStFound & \rStThird & \rPcStLab & \rStPayFound & \rStWrong \\
\rowcolor{g2}\Sys, all rules & \rTierAnyDet & \rTierAnyThird & \rTierLabAny & \rTierEvDet & \rTierAnyCross \\
\rowcolor{g2}\quad payload and host & \rTierEvDet & \rTierThirdEv & \rTierLabEv & \rTierEvDet & \rTierEvCross \\
\bottomrule
\end{tabular}}
\vspace{-3ex}
\end{table}

Our main baseline is Splunk's Enterprise Security Content Update (ESCU) rules~\cite{splunkescu}, which defenders deploy for these \cves. We run them unmodified on the same logs. No released tool compiles a \poc into detection rules for logs, so we built two versions of \sys with simpler extraction, \sysCQ and \sysEX (\secref{eval:parts}). Both differ from \sys only in extraction. \poirot and \provg search provenance graphs for an analyst's query graph. We compare \sys with them on robustness to evasion (\secref{eval:rq4}) and on \provg's benchmark (\secref{eval:darpa}). Detections by host rules, and by content or payload rules that pass the fragment test, count as \emph{evidence}.

\subsection{Coverage}
Splunk's ESCU rules for our \cves include \rRcWebRules web rules, \rRcRequestLine of which test only the request line. The web rules cover \rRcCvesWeb \cves. For \rRcSameEndpoint of these, \sys's request rule tests the same endpoint. ESCU also covers \rRlCves instrumented \cves, and we sent \rRlLegit legitimate and \rRlCrawl crawl requests to their applications. The ESCU rules and \sys's request rules flag the same \rRlRuleFlag requests.

\begin{figure}[t]
\centering
\includegraphics{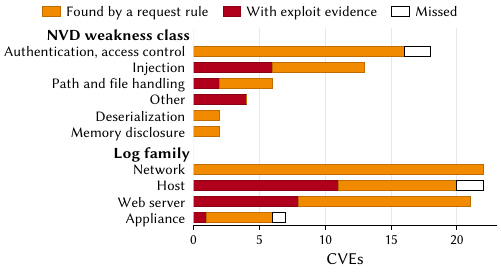}
\caption{\cves found by NVD weakness class and by the log families of their runs. A \cve whose run
has logs from several families counts under each.}
\label{fig:classes} \vspace{-3ex}
\end{figure}

Payload and host rules find \rTierEvDet of the \rCorpusTotal \cves with evidence of the exploit (Table~\ref{tab:compare}). Including request rules, \sys finds \rTierAnyDet \cves, in \rCapFound of the \rCapTotal distinct runs. The four Juniper \cves share a run, as do the two ScreenConnect and the two CrushFTP \cves. \Sys misses MOVEit's CVE-2024-5806, whose key step uses SFTP, and Outlook's CVE-2023-23397, whose \poc only sends a message. The rules find \cves in every weakness class and log family (Figure~\ref{fig:classes}). The compiled \pocs range from \rBdPocMin to \rBdPocMax lines (median \rBdPocMedian).

\subsection{Detection and False Alerts}
\label{eval:volume}
\label{eval:rq1}
\subsubsection{Splunk's \texorpdfstring{\cve}{CVE}-Specific Rules}
Splunk recorded the logs of \rCorpusSplunk \cves, and its ESCU rules include a \cve-specific rule for \rPsAvail of them. Splunk unit-tests each rule on its attack data for that \cve, and all \rPsRuleFired rules pass. We compiled \sys's rules without access to the logs. They find \rPsEncAny of these \rPsAvail \cves and all \rPsRest other \cves. No ESCU rule covers \rNoRuleCves of those \rPsRest: \rNoRuleWeb instrumented web \cves, the \rLabWin Windows \cves and CVE-2024-21683. ESCU rules outside this comparison cover the other \rRuleOther \cves (Appendix~\ref{app:setupdetail}). On the logs of our instrumented runs, \rPcSplLab of the \rRlCves rules for instrumented \cves fire. Splunk's \cve rules thus find \rPcSplFound of the \rCorpusTotal \cves in Table~\ref{tab:compare}.

\subsubsection{Other \texorpdfstring{\cves}{CVEs}' Logs}
We compare alert volume on the \rSubsetCves third-party \cves whose Splunk rule fires on the \cve's logs, so both detectors have a rule for each. These \cves form \rCcEncPairs ordered pairs that do not share a run. Splunk's rules alert on \rCcRuleFp of these pairs, with \rCcRuleAlerts alerts in total. A single rule for Citrix ShareFile raises \rCcTopRuleAlerts of these alerts. It fires on the logs of two gateway runs and alerts in \rCcTopRulePairs pairs. In the remaining \rCcOtherRulePairs pairs, the alerts come from \rCcOtherRules other rules. \Sys's rules alert on \rCcEncFp pairs. In Tables~\ref{tab:compare} and~\ref{tab:llm}, we count wrong-\cve pairs among all \rLibGridPairs ordered pairs of the \rCorpusTotal \cves from different runs. \Sys's rules alert on \rTierAnyCross of these pairs. These include the \rCcEncFp pairs above. Each of the other \rCcOutsideFp pairs includes a \cve outside the \rSubsetCves third-party \cves.

\subsubsection{Benign Traffic}
Neither Splunk's \rBhCveRules \cve-specific rules for the \rPsAvail \cves nor \sys's families alert on three public datasets: \rAtEvents host events over \rAppEpHours hours on \rAtHosts ATLASv2~\cite{riddle2023atlasv2} hosts, \rBtwoRequests requests from the NASA web server's production access log~\cite{nasahttp}, and \rSbWindows 15-minute host windows of a SOCBED~\cite{uetz2021reproducible} Windows domain. In SOCBED, \rSbHosts hosts logged \rSbRuleEvents Sysmon events, and \rSbRuleFired of Splunk's generic host rules raise \rSbRuleAlerts alerts. We also take \rExtCsicRequests normal web-store requests from CSIC 2010~\cite{csic2010} and split them into \rExtCsicWindows windows of \rExtCsicWindowSize requests. On \rLgCaps of the \rLabCaptured instrumented applications, we sent \rLgRequests legitimate requests to the vulnerable endpoint and \rHnRequests crawl requests. We also ran a sustained workload on \rLtDriven of these applications for \rLtMinutes minutes each. Its \rLtRequests requests use the vulnerable features with benign inputs. The OWASP ZAP baseline scan~\cite{zap} sent \rZpRequests crawl requests to the same \rZpLab applications and to \rZpSimilar other releases of their products. We matched the rule set against each request and each application's session (Figure~\ref{fig:alerts}).

\subsubsection{Matched Controls}
In \rCtCves of the \rApxLabFixed instrumented \cves whose \poc sends a payload fragment, the exploit injects an expression or a command. We replayed each of these \pocs with a legitimate client's value in the payload parameter. Payload rules align on all \rCtExpEv exploits and raise no alert on the controls or their \rCtReqs requests. Request rules align on both, like an analyst's rule over the request line. The sidecar records no injected command during any control.

\subsubsection{Request-Rule Leads}
When an exploit and ordinary use differ only in the body or in what happens on the target, a rule over the request line cannot tell them apart. \Sys's request rules flag \rRgEndLeadsLt of the \rRgSust requests in the sustained workload as \emph{leads} for an analyst, and \sysCQ's path rules flag the same \rRgEndLeadsLt. Across all \rRgUnits legitimate and crawl requests, endpoint rules raise \rRgEndLeads leads. Parent-directory rules add \rRgParAdd and find ColdFusion's CVE-2023-29298, and context-root rules add \rRgRootAdd and find WS\_FTP's CVE-2023-40044. In total, request rules raise \rRgAllLeads leads. Payload and host rules raise \rRgAllAlertsLt alerts on the sustained workload and \rRgAllAlertsOther elsewhere. Request rules flag nothing on \rLtSilentApps of the \rLtDriven applications under sustained load.
\begin{figure}[t]
\centering
\includegraphics{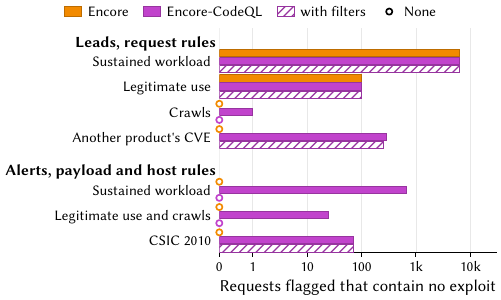}
\caption{Leads from request rules and alerts from payload or host rules on benign traffic (log scale). Circles mark zero. No family alerts on the ATLASv2, NASA and SOCBED logs.}
\label{fig:alerts} \vspace{-3ex}
\end{figure}

\subsubsection{Exchange-Level Matching}
Since every request in a session's event graph is attached to the serving process, a payload rule could
match its path in one request and its content in another. When we match exchange by exchange, the payload
rules of all \rPrPayCves instrumented \cves whose exploit sends a payload still align, matching \rPrPayExchanges of the \rPrExchanges exploit exchanges. We also mixed \rIvLegit legitimate and \rIvExploit exploit requests into the sustained workload. Payload rules align on \rIvExploitPay of the exploit requests and never on a legitimate request, and no other family alerts on any of the \rIvCves sessions.

\subsubsection{Alert Localization}
\label{eval:point}
We localize each request alert to the serving process, which also appears in host records. The kernel logged this process accepting \rAcJoined of the \rAcFlows requests relayed in the \rAcCaptures exploit, legitimate and crawl runs. The rules that find \rHostOnly \cves match only host operations (Appendix~\ref{app:localization}).

\subsection{Cross-\texorpdfstring{\cve}{CVE} Alerts}
\begin{figure}[t]
\centering
\includegraphics{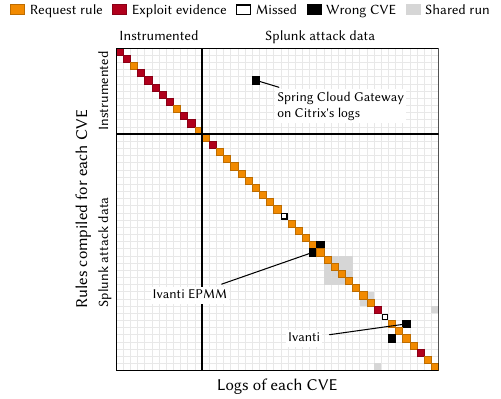}
\caption{Each \cve's rule set (rows) against each \cve's logs (columns), with our recorded runs first. A black cell off the diagonal is an alert on another \cve's logs. Shaded pairs share a run, and we leave them out.}
\label{fig:grid} \vspace{-3ex}
\end{figure}

We run all \rLibFamilies families against every \cve's logs and count alerts on \rLibGridPairs of the \rLibGridPossible ordered pairs of distinct \cves, skipping the \rLibPairsSkipped pairs that share a run. The rules of \rLibQuiet families alert on no other \cve's logs, and the rule set alerts on \rLibGridFp pairs (Figure~\ref{fig:grid}). Of these pairs, \rLibFpIvanti are between Ivanti \cves, including the \rLibFpEpmm pairs of the two Endpoint Manager Mobile (EPMM) \cves, whose exploits use the same \CodeIn{/mifs/} API. In the remaining pair, the \CodeIn{/actuator/} prefix of the Spring Cloud Gateway rule matches an \CodeIn{/actuator/} request in Citrix's logs.

\subsection{Ablations}
\label{eval:parts}
\begin{figure}[t]
\centering
\includegraphics{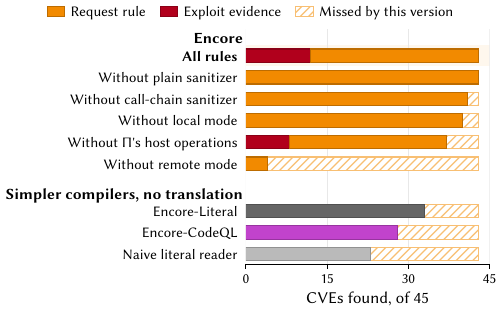}
\caption{\cves found with one part of \sys removed, and by simpler compilers without \sys's target-side translation. The full \sys finds \rTierAnyDet \cves, including every \cve found by a simpler compiler.}
\label{fig:components} \vspace{-3ex}
\end{figure}

\sysCQ uses the static analyzer CodeQL~\cite{codeql,avgustinov2016ql} and its models of the HTTP calls in requests, urllib, httpx and aiohttp to find the string literals that flow into each request. A literal in
a request's URL path becomes a \emph{path rule}, and a literal of eight or more characters elsewhere becomes
a \emph{content rule}. \sysEX uses our extractor, which resolves aliases, inlines helper functions, rebuilds composed
strings and reads the method, path, query, body and headers of every call.
It emits \rStRules rules for \rStCvesWithRules \cves. We also remove \sys's target-side translation ($R^*$, label generalization and the method map) from both. Their rules then find \rBcCqOwn (\sysCQ) and \rBcStOwn (\sysEX) of the \rCorpusTotal \cves, and a naive literal reader's rules find \rBcNvOwn. \Sys finds \rTierAnyDet (Figure~\ref{fig:components}), including every \cve found by \sysCQ or \sysEX with or without the translation. In an exact McNemar test on the \rCorpusTotal paired results,
each difference has $p \le \rMcMaxP$.

\begin{table}[t]
\tablestyle\centering
\caption{\Sys against two language models that write rules from the same \poc, scored with one matcher on the same logs. In the last row, we count payload and host alerts on the \rLgRequests legitimate-use and \rHnRequests crawl requests.}
\label{tab:llm}
\begin{tabular}{@{}l>{\columncolor{g2}}rrr@{}}
\toprule
& \Sys & Claude Opus 5 & GPT-6 Astra \\
\midrule
\cves found, of \rCorpusTotal & \rTierAnyDet & \rLlmOpusFound & \rLlmGptFound \\
Wrong-\cve pairs, of \rLibGridPairs & \rTierAnyCross & \rLlmOpusWrong & \rLlmGptWrong \\
\cves with exploit evidence & \rTierEvDet & \rLlmOpusEv & \rLlmGptEv \\
\quad wrong-\cve pairs of those rules & \rTierEvCross & \rLlmOpusEvWrong & \rLlmGptEvWrong \\
Alerts on legitimate and crawl requests & \rLlmEncBenign & \rLlmOpusBenign & \rLlmGptBenign \\
\bottomrule
\end{tabular}
\end{table}

\subsubsection{Language Models}
We gave Claude Opus 5 and GPT-6 Astra, both at high reasoning effort, the \poc compiled by \sys, a description of the target's logs, the rule format and a worked example. Each model wrote its rules once per \cve, without tools (Appendix~\ref{app:llm}). Claude Opus 5's rules find \rLlmOpusFound of the \rCorpusTotal \cves and GPT-6 Astra's find \rLlmGptFound, against \rTierAnyDet for \sys (Table~\ref{tab:llm}). The models' rules alert on another \cve's logs in \rLlmOpusWrong and \rLlmGptWrong pairs, mostly through rules that match any path. \Sys's rules do so in \rTierAnyCross pairs. Both models' payload rules also flag \rLlmAlertHits of the \rLlmAlertOf legitimate requests to GeoServer as exploits, while \sys's payload and host rules flag none of the \rLgRequests legitimate-use and \rHnRequests crawl requests.

\subsubsection{Fragment Test}
CodeQL returns every string sent by a \poc, and \sysCQ turns each one into a rule. Its content rules alert on \rCqLtContent requests of the sustained workload, and its Spring4Shell rule matches any \CodeIn{.jsp} page. Its rules also flag \rXaCqWrong legitimate requests as leads for another product's \cve and alert in all \rCqCsicUnits windows of CSIC 2010. \Sys's rules do neither (Figure~\ref{fig:alerts}). Even with \sys's filters, \sysCQ's path rules still flag \rRgCfXaWrong legitimate requests and alert in every window. Its filtered content rules find only \rCqRungEv \cves with evidence, all among the \rTierEvDet \cves found by \sys with evidence. \Sys finds the other \rCqRungGap through host operations (\rCqRungGapHost) and through payload fragments that are not among CodeQL's literals (\rCqRungGapPay).

\subsubsection{Payload and Host Rules}
A payload fragment appears in the \pocs of \rApxFixed of the \rCorpusTotal \cves. When the logs include request
bodies, payload rules find \rPayDetLab of the \rApxLabFixed instrumented \cves whose \poc sends a payload fragment.
Splunk's published web logs contain only the request line, so on those logs a payload rule matches only
fragments in the path. Payload rules find \rTierPayCves \cves with evidence, and host rules find \rTierHostCves. On third-party logs, they find Bricks (CVE-2024-25600), PrintNightmare (CVE-2021-1675) and
CrushFTP (CVE-2024-4040) with evidence. Neither kind alerts on another \cve's logs or on any of the \rRgUnits legitimate and crawl requests and their sessions.

\subsubsection{Removing Components}
If we compile each \poc in local mode only, without the target-side translation, the rules find
\rCompLocalOnly of the \rCorpusTotal \cves. Without the call-chain sanitizer, they find \rCompPlainLost
fewer \cves. Without value propagation, they find only \rAbVpDet \cves and none of them with evidence. If
we turn off inlining, the rules find \rAbInDet \cves and \rAbInEv of them with evidence. If we restrict
$\Pi$ to the \rPaKept of its \rPaIdent identifiers that map to protocol-level operations, the rules miss
every \cve found only by a host rule. The compiled families contain \rFidNamed of the \rFidSent resource requests sent by the independent \pocs of the \rFidCves instrumented web \cves. When a SIEM runs the rules as conjunctions of field
predicates, they find the same \rFieldDet \cves and alert on the same \rFieldCross wrong-\cve pairs
(Appendix~\ref{app:siem}).

\subsection{Operands in \texorpdfstring{\cve}{CVE} Descriptions}
\label{eval:rq2}
If a \cve's NVD description contained a rule's operands, the rule could be written without
exploit code. The \rRqTwoAnchors operands in our rules come from \rRqTwoCves families. Only \rRqTwoDescSup of these operands appear in the NVD descriptions, and all are product, component
or file names (\rRqTwoSupList). For \rRqTwoZero of the \rRqTwoCves \cves, the description contains none
of them.

\subsection{Performance}
\label{eval:rq3}

\begin{figure}[t]
\centering
\includegraphics{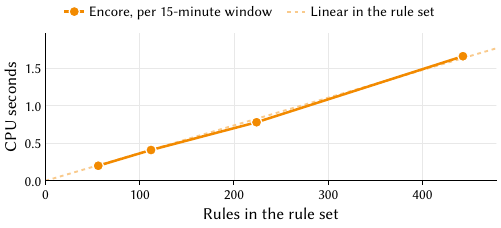}
\caption{CPU seconds to match a SOCBED client's 15-minute window with fractions of the rule set and its linear fit.}
\label{fig:scaling} \vspace{-2ex}
\end{figure}

We compile a \poc in a median of \rCompileMedian and at most \rCompileMax seconds, and we run the full
cross-\cve grid of \rCostAllAligns rule alignments in \rCostAllMin CPU-minutes (\rCostPerAlignMs ms per
alignment). In streaming mode, the \rTpMembers rules match a SOCBED client's 15-minute window of
\rTpWinEvents events using \rTpWinSec CPU-seconds and \rTpRssMb MB of memory (about \rTpEvPerSec events
per second). They match an hour's window of \rTpWinHourEvents events in \rTpWinHourSec seconds (about
\rTpWinHourEvPerSec events per second). One core can process the logs of about \rTpHostsPerCore such hosts
(\rTpHostDayMin CPU-minutes per host-day). Matching time grows linearly with the size of the rule set
(Figure~\ref{fig:scaling}).

\subsection{Generalization and Evasion}
\label{eval:rq4}
All detected exploits, including those with evidence, use code other than the compiled \poc. On Splunk's logs, the rules find \rTierAnyThird of the \rCorpusSplunk \cves in \rCapThirdFound of \rCapThird distinct runs. In \rClDiff of the \rClLogged \cves whose logs show the client, the exploit uses a different client from the compiled \poc. The rules find \rClDiffFound of them. The community templates of the open-source Nuclei scanner~\cite{nuclei} exploit \rNuReached of our targets end to end, with different payloads, commands and request order. Our rules find \rNuFoundReached of these and alert on no other \cve's logs.

\subsubsection{Alternative \texorpdfstring{\pocs}{PoCs}}
In \rAhFixedCves instrumented web \cves, the vulnerability is at a single endpoint, and we compiled \rAsPerCve additional public \pocs for each (\rAhFixedTotal alternatives). In the other two, CVE-2021-41773 and CVE-2025-24813, each author picks a different path. We took the most-starred public Python \pocs by distinct authors, excluded forks, code that shares half or more of its lines with the compiled \poc and code that sends no triggering request, and compiled the rest under the frozen $\Pi$. The closest alternative shares \rAhMaxSim of its lines with the compiled \poc. Rules compiled from \rAhFixedFound of the \rAhFixedTotal alternatives find their \cve on the same logs, between \rAsMin and \rAsMax of the \rAsPerCve alternatives for each \cve. In \rAhFixedShared of these \rAhFixedFound cases, the two rule sets test a common endpoint. Payload rules also match other authors' exploits. On the instrumented runs, they find all \rApxLabFixed instrumented \cves whose \poc sets a fragment and raise no alert on other \cves' logs or on legitimate and crawl requests. On a run with a third author's \poc for each of the \rAlcCves instrumented web \cves, the same rules find all \rAlcFound and alert on \rAlcWrong wrong-\cve pairs.
\begin{figure}[t]
\centering
\includegraphics{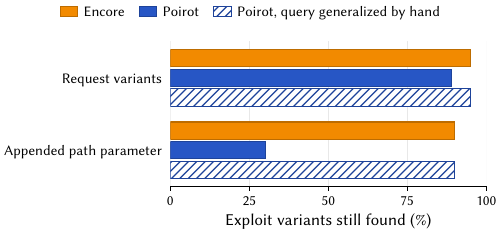}
\caption{Share of each exploit's \rEvModes variants found by each detector across the \rEvCves instrumented web \cves. We evaluate both \sys and \poirot on the \emph{request variants}. The hatched bar shows \poirot with a hand-generalized query.}
\label{fig:evasion} \vspace{-2ex}
\end{figure}

\subsubsection{Rule Evasions}
{\tolerance=2000 From the published rule evasions of AMIDES~\cite{uetz2024you} and SPECTRA~\cite{shoaib2026spectra}, we adapt \rEvModes transformations to HTTP requests. Each rewrites a request without changing the requested resource or method (Appendix~\ref{app:evasion}). We apply each to the exploits of the \rEvCves instrumented web \cves (Figure~\ref{fig:evasion}). On the original exploits, \sys finds all \rEvCves, \poirot finds \rEvPoiControls and \provg finds \rEvPgControls. \Sys finds \rEvEncKept of the \rEvEncTotal variants, and its payload rules match \rEvPayKept of the \rEvPayTotal variants with a payload. Neither \sys nor \poirot alerts on a wrong \cve under any variant, and \provg
alerts on \rEvPgWrongPairs distinct wrong-\cve pairs. With its query generalized by hand, \poirot does as
well as \sys. \provg abstracts each label to a coarse type, so no transformation affects it, and its alerts include no process.\par}
\subsection{Provenance-Graph Alignment}
\label{eval:darpa}

We also evaluate \sys's alignment on provenance graphs, against \poirot~\cite{poirot2019} on \rEncDatasets datasets from DARPA TC Engagements 3 and 5~\cite{darpae3,darpae5} and ATLASv2~\cite{riddle2023atlasv2}, and against \provg~\cite{altinisik2023provg} on the \rEncPDatasets E3 datasets (Table~\ref{t:match}). We use \rPairsPos positive and \rPairsNeg negative template--target pairs per dataset. On E3, \provg's released code samples and labels the pairs, and we train \provg with that code. On E5 and ATLASv2, we sample pairs with \provg's protocol and label them by exact subgraph search. With refinement off, \sys's learned filter discards no positive pair, and its F1 of \rMatchScrMin to \rMatchScrMax exceeds \poirot's on every dataset and \provg's on every E3 dataset. Figure~\ref{fig:pairs} in the appendix shows sample pairs.
\begin{table}[t]
\tablestyle\centering
\setlength{\tabcolsep}{3.5pt}
\caption{Alignment accuracy on \rPairsPos positive and \rPairsNeg negative template--target pairs per dataset. \provg's released code samples and labels the E3 pairs, and we label the others by exact subgraph search. In \sys rows, refinement is off.}
\label{t:match}
\begin{tabular}{@{}>{\centering\arraybackslash}m{1.75cm}lrrrrrrrr@{}}
\toprule
& System & TP & FP & FN & TN & Acc. & Prec. & Rec. & F1 \\
\midrule
\rTableMatchBody
\bottomrule
\end{tabular} \vspace{-2ex}
\end{table}

\section{Discussion and Limitations}
\label{s:discussion}

\PQ{How should defenders deploy the rules?} A request rule flags traffic to the vulnerable endpoint whether or not the target is patched, like the analysts' web rule for the \cve. If the endpoint provides a legitimate feature, request rules also flag benign requests to it (\secref{eval:rq1}), so we intend them for hunting. Payload rules match the \poc's literal fragments in logged request content, and host rules match the target's host logs. Neither kind flags any legitimate or crawl request (\secref{eval:rq1}). A team that pages an analyst on every alert can deploy them without request rules. Because we generate each family's rules at several path levels, a rule still matches when the attacker changes an untested operand, such as an uploaded file's name or the command (\secref{eval:rq4}).

\PQ{Why compile without running the \poc, and how do we keep\/ $\Pi$ current?} To run a \poc, we would need a live target with the vulnerable version for every \cve. Public \poc code is
also unvetted, and El Yadmani et al.~\cite{elyadmani2025securepoc} estimated that about 2.5\% of over
20{,}000 GitHub-hosted \poc repositories are likely malicious. \Sys analyzes each \poc without executing it. We trace each library's calls once, without a vulnerable target. $\Pi$ is keyed by operating system, host collector, runtime and library version, so after a library release we retrace only that library. We need only the \poc to add a new \cve.

\section{Related Work}
\label{s:relwk}

\PP{Provenance-based detection.}
Inam et al.~\cite{inam2022sok} survey intrusion detectors that build provenance graphs from host logs~\cite{king2003,lpm2015}. \textsc{Holmes}~\cite{holmes2019} correlates audit events through rules for
attack tactics, \textsc{Sleuth}~\cite{hossain2017sleuth} propagates tags along information flow, and
\textsc{NoDoze}~\cite{nodoze2019} ranks alerts by the rarity of their provenance paths.
Anomaly detectors such as \textsc{Unicorn}~\cite{unicorn2020}, \textsc{Kairos}~\cite{cheng2023kairos}, \textsc{Flash}~\cite{rehman2024flash}, \textsc{ProGrapher}~\cite{yang2023prographer}, \textsc{ShadeWatcher}~\cite{shadewatcher} and \textsc{Sigl}~\cite{sigl} learn from benign provenance graphs and flag deviations. \Sys uses no benign training data and names the \cve behind each match.

\PP{Rule generation.}
Detection engineers write rules one threat at a
time in Sigma~\cite{sigmasignatures}, SIEM query languages~\cite{splunk,uetz2024you} or network signature
languages such as Snort~\cite{snort}.
\poirot~\cite{poirot2019} aligns an analyst-written query graph with a provenance graph and scores partial alignments by information flow. \provg~\cite{altinisik2023provg} uses order embeddings~\cite{vendrov2015order,lou2020neural} to find a query graph inside large provenance graphs. \textsc{Extractor}~\cite{satvat2021extractor} and \textsc{ThreatRaptor}~\cite{gao2021enabling} derive such queries from threat reports. LLMCloudHunter~\cite{schwartz2025llmcloudhunter} and Cai et al.~\cite{cai2026sigmerge} prompt language models with threat reports to generate Sigma rules for cloud API and host process logs. Several systems derive network intrusion detection rules from exploits. MetaSymploit~\cite{wang2013metasymploit} symbolically executes Metasploit scripts, and Kobayashi et al.~\cite{kobayashi2023idssig} explore the execution paths of \poc code. RuleXploit~\cite{papoutsis2025rulexploit} prompts a model with an exploit and its \cve description to write rules, which it validates by running the exploit. Yamamoto and Katsurai~\cite{yamamoto2026verification} synthesize rules from recorded HTTP attack requests. \Sys compiles request, payload and host rules from the \poc's operations on the target without running the exploit or a model.

\PP{Web attack detection.}
Web anomaly detectors model the parameters, lengths and character distributions of an application's requests and flag deviating requests~\cite{kruegel2003anomaly,song2009spectrogram}. Web
application firewalls match requests against generic attack rules such as the OWASP Core Rule
Set~\cite{owaspcrs}. Both detect classes of attack such as injection, and each \sys rule targets one disclosed \cve.

\PP{Vulnerability-driven signatures.}
Shield~\cite{shield2004}, Brumley et al.~\cite{brumley2006}, ShieldGen~\cite{shieldgen2007}, Vigilante~\cite{vigilante2005} and Bouncer~\cite{bouncer2007} analyze the vulnerable program to generalize from one exploit to the class of inputs that exploit the vulnerability. \Sys does not need that program.

\section{Conclusion}
\label{s:conclusion}

A public \poc is available weeks before any rule for its \cve, and its code specifies the exploit's requests, payloads and host operations. \Sys statically compiles that code into the \cve-specific hunting rules otherwise written by hand. It needs neither a vulnerable target nor exploit logs, and its rules run on web-server, proxy and host logs. Its rules detect exploits of \rTierAnyDet of the \rCorpusTotal \cves, and its payload and host rules detect \rTierEvDet without alerting on legitimate traffic. $\Pi$ maps every log-relevant call in the Python \pocs of \rEcoKevMapped KEV-listed \cves. We plan to compile rules for these \cves.

\bibliographystyle{ACM-Reference-Format}
\bibliography{bib-files/hassan-misc,bib-files/hassan-prov}

\appendix
\section*{Appendix}

\section{Ethics}
\label{app:ethics}

We ran every exploit in this paper in an isolated container or a virtual machine against vulnerable software deployed in our lab. We targeted no third-party system. Third-party logs and benign traffic come from public datasets released for research. Every compiled \poc is public. No human subjects were involved.

\section{Open Science}
\label{app:openscience}

Upon acceptance, we will open-source the \poc compiler, the alignment engine, the evaluated \pocs and compiled templates, the code that runs Splunk's deployed rules on our logs, the scripts that run every experiment with our outputs, and download scripts for public datasets.

\section{Compiling a \texorpdfstring{\poc}{PoC} into a Template}

\subsection{Constructing the API-to-Operation Map}
\label{app:pi}

Each operation record of $\Pi$ specifies how the compiler lowers the call (an HTTP exchange, a file operation, a process creation or a socket operation), the resulting event class in $\Sigma_\sigma$, and the argument that contains its operand. For each API, we run a no-op script and a script that calls the API under the host collector, normalize both event streams and take their multiset difference. We repeat this for several argument values and program states and keep the operations present in every difference.

\subsection{Template Extraction}
\label{app:algo}

The compiler enumerates the execution paths through a \poc and resolves the target endpoint of each. It lowers each path in both run modes and emits every mapped call in program order.

\subsection{Process Scoping and Path Enumeration}
\label{app:designdetail}

Before extraction, a sanitizer normalizes the \poc. It rewrites path-library and HTTP-session idioms into plain calls, substitutes command-line defaults, collapses exception handlers and keeps only statements needed by mapped calls. In its \emph{call-chain} mode, the sanitizer keeps calls to functions defined in the \poc when their call chain leads to an anchor, and the compiler inlines those functions at their call sites. In its \emph{plain} mode, it keeps only the function bodies. The compiler splits the syntax tree at every branch into straight-line programs, up to 256 per \poc. The compiler assigns program-scope calls to the \poc's process. A process-creating call introduces a new subject for the later calls in the created process. The distinct templates of both sanitizer modes and both run modes form a family. The compiler merges textually identical rules and removes rules without an edge. If it cannot resolve a request URL built by the code, it takes a URL from the \poc's usage text whose host is a placeholder, a loopback or a private address.

The compiler selects at most two literal fragments per \poc for payload rules, from the body, form,
JSON, query or non-standard header arguments of its HTTP calls. A fragment passes the \emph{fragment test} if it is at least eight characters long and contains a path traversal or one of \CodeIn{\{\}()\$;`}. Markup brackets do not count toward the test. Format placeholders in a fragment become wildcards. The compiler adds one payload variant of each request rule per fragment.

\subsection{Emission Order}
\label{app:compdetail}

The compiler emits edges in program order along each straight-line path, and that order is the
template's $\prec$. When logs contain no timestamps, we impose neither the order nor the window condition.

\subsection{Release, Freeze and Template Details}
\label{app:freeze}
We will release $\Pi$, $\Sigma_\sigma$, $T$, $L$, $R^*$ (\secref{s:design:transfer}), the label normalization and the collector profiles (Table~\ref{tab:collectors}) under a commit ID and the SHA-256 hash of a manifest. If two fields embed the same variable in different text, as \CodeIn{"/user/"+x} and \CodeIn{"id="+x} do, their values differ for every $x$, and the fields share no symbolic variable. Neither do fields whose selectors are encoded differently. We replace each such field with a fresh anonymous wildcard that matches any value. We represent each vertex-label component as a normalized concrete value, a symbolic token shared by linked fields or a fresh anonymous token. A call is hidden from the compiler if its target is not syntactically fixed or if its call site or operands depend on monkey patching, reflection, dynamic imports, run-time callbacks or custom raw-socket clients. The compiler accepts every parsable \poc. Across the
compiled \poc corpus, templates contain a median of 6 edges and 4 concrete vertices, and 93\% have more
than one edge.

\subsection{Observability}
\label{app:observe}
We call an operation or vertex \emph{structurally} observable if a deployed collector logs it, and a label component \emph{field} observable if the log format includes it. We match each rule as a whole, and only on targets whose collectors record every required role and event class.

\section{Aligning Templates to Logs}

\subsection{Seeding Candidates}
\label{s:align}

A candidate is the $h$-hop neighborhood of a process vertex in \evtg. We try candidates in turn and raise the alert at the first candidate with a valid mapping of the template.

\subsection{Learned Candidate Filter}
\label{app:screen}

On large provenance graphs, \sys prunes candidates with a learned filter before refinement. The filter discards a candidate when the order violation $\mathrm{viol}(z_{tem},z_{cand})=\|\max(0,z_{tem}-z_{cand})\|_2^2$ between the template's encoding $z_{tem}$ and the candidate's $z_{cand}$ exceeds a threshold $s$. The encoder embeds each vertex from its type and hashed label tokens, passes messages over three layers with non-negative weights per event class and direction, and appends hashed counts of the vertex labels and of the labelled paths of up to two hops from the root. We train it with the order-embedding loss~\cite{vendrov2015order,lou2020neural} on \rDarpaPoolPairs template--candidate pairs from the earliest part of the E3 CADETS time span, choose $s$, $h$ and $k$ on validation pairs from the following part, and freeze them for every test set. On these benchmarks, the candidates are the $h$-hop neighborhoods of the target vertices labelled with the template's rarest label. With the filter, \sys discards up to \rScreenPruneMax\% of candidates and finds every template's true match across the full E3 test graphs.
We compare the filter's accuracy with \provg and \poirot in \secref{eval:darpa}.

\subsection{Timing and Staging}
\label{app:timing}
The emission order $\prec$ (Appendix~\ref{app:compdetail}) is a linear extension of the edges' dependency constraints. We timestamp a matched template edge with $t(\mu_E(e))$, the time of its last event. If the match passes through intermediate process creations, the last event is on the edge that ends at the target vertex. We allow a clock skew of up to $\epsilon$ between collectors on the same target. During refinement, we require $t(\mu_E(e_i)) \le t(\mu_E(e_{i+1})) + \epsilon$ for consecutive edges in $\prec$, and $\max t - \min t \le \Delta$ over matched edges. Every edge of $P_e$ before $\mu_E(e)$ is a process creation by the subject or its descendants. On edges of some classes, the path contains no intermediate creation. For an edge \emph{into} the subject, such as an inbound protocol record, $P_e = \mu_E(e)$. For a process-creation edge, the prefix is a chain of creations, so an ancestor may spawn the child through intermediate shells. The bound $k$ lets a match pass through process creations recorded in host logs between a \poc's operations. Distinct template edges must end in distinct events. No compiled rule is all wildcards.

\subsection{Refinement and Labels}
\label{app:aligndetail}

The matcher extends an injective mapping outward from the seed process. The matcher accepts a candidate label if it matches a disjunct of the template label. A placeholder for a command-line argument or a payload matches any label. A disjunct with a wildcard compiles to a pattern, and before matching it we normalize path separators and, on Windows paths, case. Any other disjunct must equal the label under the same normalization.

\section{Evaluation Details}

\definecolor{pcevink}{HTML}{B1001C}
\definecolor{pcreqink}{HTML}{B34400}
\DeclareRobustCommand{\pglyph}[1]{\tikz[baseline=-0.55ex]{#1}}
\DeclareRobustCommand{\phid}[1]{\makebox[0pt][l]{\textpdfrender{TextRenderingMode=Invisible}{#1}}}
\DeclareRobustCommand{\psrcS}{\phid{S}\pglyph{\fill (-2.2pt,-2.2pt) rectangle (2.2pt,2.2pt);}}
\DeclareRobustCommand{\psrcL}{\phid{I}\pglyph{\draw[line width=0.7pt] (-2pt,-2pt) rectangle (2pt,2pt);}}
\DeclareRobustCommand{\pfP}{\textbf{\textcolor{pcevink}{P}}}
\DeclareRobustCommand{\pfH}{\textbf{\textcolor{pcevink}{H}}}
\DeclareRobustCommand{\pfR}{\textbf{\textcolor{pcreqink}{R}}}
\DeclareRobustCommand{\pfN}{}
\DeclareRobustCommand{\pcqY}{\phid{Y}\pglyph{\fill (0,0) circle (2.3pt);}}
\DeclareRobustCommand{\pcqN}{}
\DeclareRobustCommand{\prF}{\phid{F}\pglyph{\fill (0,0) circle (2.3pt);}}
\DeclareRobustCommand{\prS}{\phid{S}\pglyph{\draw[line width=0.7pt] (0,0) circle (2.1pt);}}
\DeclareRobustCommand{\prN}{}
\newcommand{\pcve}[1]{\rotatebox{90}{#1}}
\newcommand{\ptext}[1]{#1}
\begin{table}[t]
\centering
\scriptsize
\setlength{\tabcolsep}{2.1pt}
\caption{Per-\cve results. \emph{Source}: \psrcS{} Splunk's attack data, \psrcL{} our run. \emph{Found by}: \pfP{} payload, else \pfH{} host, else \pfR{} request rule. \emph{\sysCQ}: \pcqY{} finds it. \emph{Splunk rule}: \prF{} fires, \prS{} silent. \emph{Text}: share of rule operands in the NVD description. Blank: none.}
\label{t:percve}
\begin{tabular}{@{}lccccccccccccccc@{}}
\toprule
 & \pcve{2019-16116} & \pcve{2020-6287} & \pcve{2021-1675} & \pcve{2021-3129} & \pcve{2021-22205} & \pcve{2021-41773} & \pcve{2022-1388} & \pcve{2022-22947} & \pcve{2022-22963} & \pcve{2022-22965} & \pcve{2022-40684} & \pcve{2022-42889} & \pcve{2022-46169} & \pcve{2023-3519} & \pcve{2023-4966} \\
\midrule
Source & \psrcL & \psrcS & \psrcS & \psrcL & \psrcL & \psrcL & \psrcS & \psrcL & \psrcL & \psrcS & \psrcS & \psrcS & \psrcL & \psrcS & \psrcS \\
Found by & \pfH & \pfR & \pfH & \pfP & \pfR & \pfP & \pfR & \pfP & \pfP & \pfR & \pfR & \pfR & \pfP & \pfR & \pfR \\
\sysCQ & \pcqN & \pcqN & \pcqN & \pcqY & \pcqY & \pcqN & \pcqY & \pcqY & \pcqY & \pcqY & \pcqY & \pcqY & \pcqY & \pcqN & \pcqY \\
Splunk rule & \prN & \prN & \prF & \prN & \prN & \prN & \prF & \prN & \prF & \prF & \prF & \prF & \prN & \prF & \prF \\
Text & \ptext{0/2} & \pfN & \ptext{0/5} & \ptext{0/3} & \ptext{0/3} & \ptext{0/2} & \ptext{0/3} & \ptext{1/4} & \ptext{0/1} & \ptext{0/3} & \ptext{0/5} & \ptext{0/2} & \ptext{1/1} & \ptext{0/2} & \ptext{0/3} \\
\bottomrule
\end{tabular}

\medskip

\begin{tabular}{@{}lccccccccccccccc@{}}
\toprule
 & \pcve{2023-20198} & \pcve{2023-22515} & \pcve{2023-22527} & \pcve{2023-23397} & \pcve{2023-24489} & \pcve{2023-29298} & \pcve{2023-29357} & \pcve{2023-35078} & \pcve{2023-35082} & \pcve{2023-36844} & \pcve{2023-36845} & \pcve{2023-36846} & \pcve{2023-36847} & \pcve{2023-38646} & \pcve{2023-40044} \\
\midrule
Source & \psrcS & \psrcS & \psrcS & \psrcS & \psrcS & \psrcS & \psrcS & \psrcS & \psrcS & \psrcS & \psrcS & \psrcS & \psrcS & \psrcL & \psrcS \\
Found by & \pfR & \pfR & \pfR & \pfN & \pfR & \pfR & \pfR & \pfR & \pfR & \pfR & \pfR & \pfR & \pfR & \pfP & \pfR \\
\sysCQ & \pcqY & \pcqY & \pcqY & \pcqN & \pcqY & \pcqN & \pcqY & \pcqY & \pcqY & \pcqY & \pcqY & \pcqY & \pcqY & \pcqY & \pcqN \\
Splunk rule & \prF & \prF & \prF & \prF & \prF & \prF & \prF & \prF & \prF & \prF & \prF & \prF & \prF & \prN & \prF \\
Text & \ptext{0/3} & \ptext{0/5} & \ptext{1/3} & \pfN & \ptext{0/8} & \ptext{0/3} & \ptext{0/4} & \ptext{0/4} & \ptext{0/2} & \ptext{0/1} & \ptext{0/1} & \ptext{0/1} & \ptext{0/1} & \ptext{0/2} & \ptext{0/3} \\
\bottomrule
\end{tabular}

\medskip

\begin{tabular}{@{}lccccccccccccccc@{}}
\toprule
 & \pcve{2023-42793} & \pcve{2024-1708} & \pcve{2024-1709} & \pcve{2024-4040} & \pcve{2024-5806} & \pcve{2024-7593} & \pcve{2024-21683} & \pcve{2024-21887} & \pcve{2024-23897} & \pcve{2024-25600} & \pcve{2024-36401} & \pcve{2025-5777} & \pcve{2025-8088} & \pcve{2025-24813} & \pcve{2025-31161} \\
\midrule
Source & \psrcL & \psrcS & \psrcS & \psrcS & \psrcS & \psrcS & \psrcS & \psrcS & \psrcS & \psrcS & \psrcL & \psrcS & \psrcL & \psrcL & \psrcS \\
Found by & \pfR & \pfR & \pfR & \pfH & \pfN & \pfR & \pfR & \pfR & \pfR & \pfP & \pfP & \pfR & \pfH & \pfR & \pfR \\
\sysCQ & \pcqY & \pcqY & \pcqY & \pcqY & \pcqN & \pcqY & \pcqY & \pcqN & \pcqY & \pcqY & \pcqY & \pcqY & \pcqN & \pcqY & \pcqY \\
Splunk rule & \prS & \prF & \prF & \prF & \prF & \prF & \prN & \prN & \prF & \prF & \prN & \prF & \prN & \prF & \prF \\
Text & \ptext{0/5} & \ptext{0/11} & \ptext{0/2} & \ptext{0/8} & \ptext{0/1} & \ptext{0/4} & \ptext{0/6} & \pfN & \ptext{1/1} & \ptext{0/3} & \ptext{1/2} & \ptext{0/1} & \pfN & \ptext{0/2} & \ptext{0/2} \\
\bottomrule
\end{tabular}
 
\end{table}

\subsection{Per-\texorpdfstring{\cve}{CVE} Results}
\label{app:rq1detail}

Table~\ref{t:percve} lists the per-\cve results. We exploit the \rLabWin Windows \cves in a Windows 10
virtual machine with Process Monitor and Sysmon, against CompleteFTP 8.3.3 and WinRAR 7.12. The CompleteFTP \poc reads the credentials logged during first-run setup and opens the manager as administrator. In the WinRAR run, we record the exploit archive from its creation to its extraction, which writes the payload into the Startup folder.

\begin{figure*}[t]
\centering
\includegraphics{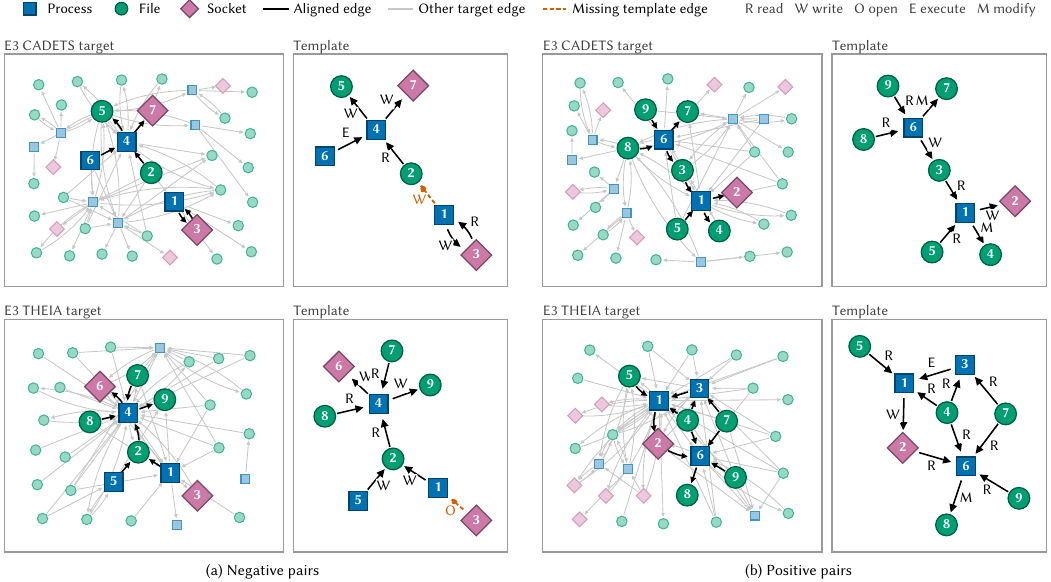}
\caption{Sample (a) negative and (b) positive template--target pairs from \provg's benchmark. \provg classifies both negative pairs as matches, and \sys classifies all four pairs correctly. \provg decides from learned embeddings, while \sys checks every template edge against the target and rejects each pair in (a), where the dashed template edge is missing.}
\label{fig:pairs} 
\end{figure*}

{\tolerance=2000 For PrintNightmare, the \poc calls \CodeIn{par.}\allowbreak\CodeIn{hRpcAsync}\allowbreak\CodeIn{AddPrinterDriver} with a driver container and sets its configuration file to \CodeIn{winhttp.dll}, then \CodeIn{kernelbase.dll}, then the \poc's DLL in each \CodeIn{old} slot of the spooler's x64 driver directory. MS-PAR (the Print System Asynchronous Remote Protocol) specifies that the server copies the files named in a container, so we wrote $\Pi$'s entry for that call as a \textsc{write} of those paths. The host rule matches any process that writes one of those paths, and in the logs it matches \CodeIn{spoolsv.exe} writing \CodeIn{evil.dll} into those slots.\par}

\subsection{Coverage over the \texorpdfstring{\poc}{PoC} Population}
\label{app:rq4detail}

The KEV catalog lists \rEcoKevWindow \cves disclosed between 1 January 2019 and our 30 September 2025 snapshot. Of these, \rEcoKevPoc have a public \poc and \rEcoKevPython have a Python \poc. Our compiler parses \rEcoKevParses of the Python \pocs, and $\Pi$ maps every log-relevant call in \rEcoKevMapped of them. Compiling the \rCorpusTotal \pocs again yields byte-identical templates.

\subsection{Setup Details}
\label{app:setupdetail}
For each \cve, two of us confirmed that different people wrote the recorded and the compiled \poc, that their repositories share no fork or upstream, and that the recorded \poc copies no payload from the disclosure's write-up. We spent 5 person-hours per \cve on average on deployment, verification and recording. ESCU rules outside the comparison in \secref{eval:rq1} cover the remaining \rRlCves instrumented web \cves, CVE-2024-21887 and CVE-2020-6287. Splunk files the logs of the last two under CVE-2023-35081 and CVE-2025-31324, and the rule for CVE-2025-31324 fires on the logs of CVE-2020-6287. The rules of \rAppEpFam families apply to the ATLASv2 host events, and those of \rAppWebFam families apply to the NASA web log. The instrumented runs contain \rLabReqTotal requests and a median of \rLabHostMedian host events per \cve. The logs of each run show the exploit's effect. ATLAS~\cite{atlas} was evaluated on ten reconstructed APT attacks that exploit six distinct \cves~\cite{riddle2023atlasv2}. LLMCloudHunter~\cite{schwartz2025llmcloudhunter}, a Sigma rule generator, was evaluated on 20 annotated cloud threat reports.

\subsection{Alert Localization and Host-Only Rules}
\label{app:localization}
Without using the matcher, we joined the proxy's side of each connection to the matching accept call on the server. Up to \rAcMaxWorkers workers shared a listening socket. In \rAcCmdCves runs, the exploit executes a command. All \rAcCmds processes started by the exploit descend from the serving process and began while that process handled the exploit request. For CVE-2022-22947, the rule compiled from the \poc contains a write and a
read of the same route under \CodeIn{*/actuator/gateway/routes/*}. The logs show the two operations as two requests, a \CodeIn{POST} that creates the route and a \CodeIn{GET} that reads it back. Both requests are attached to the serving process, which spawns the attacker's command between them, and the alignment matches both there. The host-only rules match the
driver DLL that the print spooler writes into System32 under PrintNightmare, the process chain of the
CompleteFTP local privilege escalation, and the alternate-data-stream write and \CodeIn{rar.exe} process
of WinRAR's CVE-2025-8088, whose exploit is an archive with a traversal path in an alternate data stream.
For each host-rule alert, we record the machine that ran its operations. In Table~\ref{t:percve}, both a host rule and a request rule find CrushFTP's CVE-2024-4040.

\subsection{SIEM Field Predicates}
\label{app:siem}
Defenders can run each rule as a SIEM correlation search with one field predicate per edge. We compared this form with the alignment on \rLocCves \cves. The field predicates return every logged event that satisfies any one of them, a median of \rLocFieldEventsMed events per \cve and up to \rLocFieldEventsMax. For \rLocFieldProcsOverOne \cves, these events come from several processes. The alignment returns one connected match at the serving process, with a median of \rLocWitEventsMed events. We also drew queries from \rJfDarpa datasets of DARPA Engagements 3 and 5~\cite{darpae3,darpae5} and from the endpoint detection and response (EDR) records of ATLASv2, in which every process instance is one vertex. Each query is a small subgraph of one process's events, and we paired it with a candidate that lacks the query. Of these \rJfNeg candidates, \rJfOperandsPct\% contain every operand of their query, and \rJfFieldsAccept contain all of its operations somewhere. Among \rJfHardNeg candidates with queries from other processes, \rJfHardAccept have all query operations. The predicates alert on all of these when run independently and reject them when joined on shared vertices.

\subsection{Rule Evasions}
\label{app:evasion}
AMIDES collects hand-crafted evasions of SIEM rules from enterprise networks, and SPECTRA generates families of transformations against rule syntax. The \rEvModes transformations change header case, add junk parameters, reorder parameters, percent-encode parameter values and path segments, insert whitespace and comments, write a command as an absolute path, insert \CodeIn{/./} segments, append a path parameter and rename an XML namespace. We give \poirot a query graph of the same shape, labelled with the concrete values recorded in the unmodified run. Because \poirot finds no candidate alignment for its missed variants, it computes no score for them, and its counts are the same for every score threshold $C_{thr}$ from \rEvCthrLo to \rEvCthrHi. \provg's results are the same whether we anchor its query at the highest-degree process or at the serving process.

\subsection{Language-Model Prompt}
\label{app:llm}
We described the target's logs as web-server access logs, proxy logs with request bodies, and host process and file logs. We asked for rules for a remote run, a local run or both, excluding the attacker's machine. The worked example is a rule for a \cve outside the corpus. As with \sys, the models write rules before any exploit log exists. The full prompt for each \cve and the models' outputs will be in our artifact (Appendix~\ref{app:openscience}).

\end{document}